\documentclass[10pt,twocolumn,twoside]{IEEEtran}
\makeatletter
\newif\if@restonecol
\makeatother

\usepackage[ruled,vlined,boxed,ruled,commentsnumbered]{algorithm2e}
\usepackage[dvips]{color}
\usepackage{color}
\usepackage{epsf}
\usepackage{times}
\usepackage{epsfig}
\usepackage{graphicx}
\usepackage{url}
\usepackage{amsmath}
\usepackage{amssymb}
\usepackage{amsxtra}
\usepackage{algorithm}
\usepackage{algorithmic}
\usepackage{cite}
\usepackage{listings}
\usepackage{verbatim}
\usepackage{pseudocode}
\usepackage{mathrsfs}
\usepackage{float}
\usepackage{stfloats}
\usepackage{subfig}
\usepackage{epstopdf}
\usepackage{multirow}
\usepackage{diagbox}
\newtheorem{condition}{Condition}
\newtheorem{assumption}{Assumption}

\newtheorem{theorem}{\bf Theorem}

\newtheorem{lemma}{Lemma}[section]

\newtheorem{definition}{\bf Definition}
\newtheorem{remark}{Remark}

\begin{document}
	\title{Optimal Power Control for DoS Attack over Fading Channel: A Game-Theoretic Approach}
	\author{Jie~Wang, Jiahu~Qin,~\IEEEmembership{Senior~Member,~IEEE,}
		~Menglin~Li, and Yang~Shi,~\IEEEmembership{Fellow,~IEEE}
		\thanks{J. Wang, J. Qin, and M. Li are with the Department of Automation, University
			of Science and Technology of China, Hefei 230027, China (e-mail:
			wj1993@mail.ustc.edu.cn; jhqin@ustc.edu.cn; lml95@mail.ustc.edu.cn).
			
			Yang Shi is with the Department of Mechanical Engineering, University of
			Victoria, Victoria, BC V8W 3P6, Canada (e-mail: yshi@uvic.ca).}
	}
	\maketitle
	\begin{abstract}
		In this paper, we investigate remote state estimation against an intelligent denial-of-service (DoS) attack over a vulnerable wireless network whose channel undergoes attenuation and distortion caused by fading. We use the sensor to observe system states and transmit its local state estimates to the remote center.
		Meanwhile, the attacker injects a jamming signal to destroy the packet accepted by the remote center and causes the performance degradation.
		Most of the existing works are built on a time-invariant channel state information (CSI) model in which the channel fading is stationary.
		However, the wireless communication  channels are more prone to dynamic changes.
		To capture this time-variant property in the channel quality of the real-world networks, we study the fading channel network whose channel model is characterized by a generalized finite-state Markov chain.
		With the goals of  two players in infinite-time horizon, we describe the conflicting characteristic between the attacker and the sensor with a general-sum stochastic game.
		Moreover, the  Q-learning techniques are applied to obtain an optimal strategy pair at a Nash equilibrium. Also the monotone structure of the optimal stationary strategy is constructed under  a sufficient condition.
		Besides, when channel gain is known a \textit{priori}, except for the full Channel State Information (CSI), we also investigate the partial CSI, where Bayesian games are employed. Based on the player's own channel information and the belief on the channel distribution of other players, the energy strategy at a Nash equilibrium is obtained.
	\end{abstract}
	
	\section{Introduction}
	Cyber-physical systems (CPSs) tightly integrate computation, communication and control with cyber elements and physical processes\cite{Johansson2014}. 
	With a promising future, CPSs have been applied in a large scope of infrastructures including internet of things, environmental monitoring, self-driving cars, smart grids, mine monitoring, etc\cite{Hovareshti2007,Poovendran2012}. In most of cyber-physical infrastructures, safety is a crucial problem. Whereas, due to the nature of high openness in cyber communication networks, CPSs are vulnerable to the malicious attacks from the outside.  As a result, there is an urgent need to address the issues of \textit{cyber-security}\cite{Cardenas2008,Fawzi2014,Cardenas2018a}.
	
	Researchers mainly investigated cyber-security under two types of attacks on CPS: deception attacks \cite{Ding2018} and denial-of-service (DoS) attacks
	\cite{Chen2019,Befekadu2015,Peris2015,Qin2018,Chen2019a}.
	The deception attack mainly degenerates the system performance by maliciously modifying the communication data packets, while DoS attack jams the communication channel to compromise the availability of  data packets. In this work, we are committed to remote state estimation against DoS attacks which are common and easy to implement. The adversary deploys a DoS attack which jams the communication channel to diminish the network's capacity to transmit the signals sent by the sensor.
	In a wireless communication network one has to take some factors into account such as background noise, interference, channel fading,  etc \cite{Leong2011}. The concept of signal-to-interference-plus-noise ratio (SINR) creates a model to consider these factors \cite{Adibi2007,Ding2017,Qin2020}.
	The existing communication theory reveals that the packet dropout rate is connected with the energy.
	
	Several studies investigating DoS attacks have been implemented with the standpoint of only one side, see \cite{Zhang2015,Zhang20181} for more details and the reference therein. Zhang \textit{et al}. \cite{Zhang2015} considered how to launch DoS attacks intelligently with the power consumption as little as possible in  the standpoint of attacker. The sensor should adopt rational action (by choosing to send different transmission energy level) to avoid DoS attacks, at the same time the attacker would identify sensor's action, and revise its attack method accordingly.
	If an interaction between the attacker and the sensor is considered, the defensive/offensive strategy designs will be complicated.
	The game-theoretic framework has been adopted to model the interactive actions  between  attackers and sensors  in \cite{Agah2004,Li2016}. Li \textit{et al}. \cite{Li2016} studied a scenario where a sensor sends signal to the remote center by a wireless transmission channel and a DoS attacker whose target is to deteriorate the system estimation performance obstructs the channel by using as little energy as possible. By proposing a two-player zero-sum game in which  sensors and  attackers, both have limited power, obtain their optimal solution to maximize their reward functions, \cite{Li2016} showed that the optimal solutions for sensors and attackers form a mixed strategy Nash equilibrium. Liu \textit{et al}. \cite{Liu2019} investigated the infinite-time targets of the sensor and attacker with asymmetric information over SINR network and modeled the conflicting nature between them by a Stackelberg game.
	
	However, the existing  works are built on a time-invariant channel state information (CSI) scenario in which the channel fading is stationary over the whole time horizon. Unlike wired communications that take place over a relative stable medium, the wireless transmission medium varies strongly over time.
	
	We will examine how the sensor and the attacker use the information about the channel quality (referred to \textit{Channel State Information} (CSI))
	to adjust their transmission parameters. However, the consequence of using CSI under self-interested behavior requires sophisticated game-theoretic analysis. In order to represent the channel fading statistics, a large number of samples are needed in simulation to obtain the statistically matched channel gain\cite{Sadeghi2008}, so high computational cost is required. Thus, finite state channel model is adopted \cite{Leong2011} so as to reduce the complexity.
	
	Besides, most previous works assume that the knowledge such as the channel state information (CSI) \cite{Caire1999} about other devices is available to all devices. However, this is not very possible in a real scenario. In light of this, the original general-sum stochastic game framework is changed into a static Bayesian game. In this scenario, both the sensor and the attacker acquire incomplete information, in other words, the sensor has its own channel gain, but does not acquire the attacker's. Similar things happen to the attacker.
	
	In this work, to capture the influence of the time variation in the channel quality of practical networks and obtain how different channel states affect the estimation performance,
	a  general-sum stochastic game is established to describe the interactive action between the transmitted energy for sensors and the interfering energy for  attackers over a time-varying channel state information scenario which embeds the work in \cite{Li2016} as a special case. For the incomplete information scenario, the previous game framework is changed to a static Bayesian game.
	The main contributions of our work are as follows.
	\begin{itemize}
		\item[1)] The interaction between the transmission strategy on sensors and attackers is investigated in the scenario of time-varying fading channel which embeds the previous work investigating the stationary fading channel \cite{Li2016} as a special case.  A general-sum stochastic game framework is developed to find an energy-efficient action for the sensor and the attacker simultaneously.
			The introduction of time varying channel state incurs significant technical challenges as the channel gain changes at the next moment is uncertain and the analysis is considered  in a probabilistic sense.

		\item[2)] We present a modified Q-learning method, which is called Nash Q-learning algorithm, to find the optimal solutions.  We also provide  the convergence analysis of Nash Q-learning algorithm for our stochastic game.  And  the monotone structure of the stationary Nash equilibrium strategy is constructed under a sufficient condition.
		\item[3)] In the case of known channel fading, in addition to the complete CSI scenario, partial CSI cases or incomplete information games are also considered. In this case, each participant fully acquires his own channel gain, but only knows the statistical channel information of his opponent. In this case, we study Bayesian equilibria, in which the ``optimal" energy scheme in each participant relies only on its own channel information.
	\end{itemize}
	The rest of  paper is organized as follows. We first provide the system dynamic, the method of state estimation and the fading channel model in section \ref{sec:problem setup}. The  problem of interest is then developed. The framework of the sensor-attacker game and the existence of the equilibrium strategy is demonstrated in section~\ref{sec:stochastic-game}. Section \ref{sec:stochastic-game} also provides
	the practical implementation of obtaining the optimal policy and the convergence analysis of proposed Nash Q-learning algorithm in our game. Besides, the monotone structure of the optimal stationary  strategy is also constructed. In section \ref{sec:incomplete-information-game}, the framework of incomplete information stochastic Bayesian game is formulated and the type-contingent strategy is obtained. The simulations and conclusions are given in Sections \ref{sec:numerical-examples} and \ref{sec:conclusion}, respectively.
	
	\textit{Notations:}
	Denote by $Z$ the set of non-negative integers.  $R^{n}$ represents the $n$-dimensional Euclidean
	spaces.
	For a matrix $X$, $X'$ denotes its transpose. $Tr[\cdot]$ denotes the trace of a matrix. $X\geq Y$ if $X-Y \in S^{n}_{+}$. $E[\cdot]$ denotes the expectation of a random
	variable. 
	$\delta_{ij}$ is the Dirac delta function, i.e., $\delta_{ij}=1$ when $i=j$, and $\delta_{ij}=0$ if $i\neq j$.
	
	\section{Problem Setup}\label{sec:problem setup}
\begin{figure}
		\centering
		\includegraphics[width=0.49\textwidth]{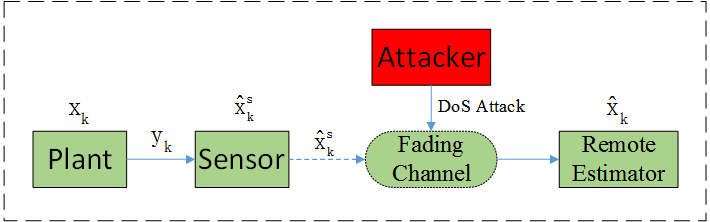}\\
		\caption{System architecture: The transmission channel is destroyed by a malicious attacker.}\label{fig:architecture}
	\end{figure}

	Consider a general discrete linear time-invariant (LTI) system (see Fig.~\ref{fig:architecture}) as follows
	\begin{align}
	x_{k+1}&=Ax_{k}+\omega_{k},\\
	y_{k}&=Cx_{k}+\nu_{k},
	\end{align}
	where $k\in\mathbb{N}$, $x_{k}\in\mathbb{R}^{n_{x}}$ represents the system state, $\omega_{k}\in\mathbb{R}^{n_{x}}$ represents the system noise, $y_{k}\in\mathbb{R}^{n_{y}}$ stands for the observation made by the sensor, and $\nu_{k}\in\mathbb{R}^{n_{y}}$ is the observation noise. Furthermore, $\omega_{k}$ and $\nu_{k}$ represent zero-mean i.i.d Gaussian noised with $E[\omega_{k}\omega_{j}^{'}]=\delta_{kj}Q$ $ (Q\geq 0)$, $E[\nu_{k}\nu_{j}^{'}]=\delta_{kj}R$ $(R>0)$, $E[\omega_{k}\nu_{j}^{'}]=0$, $\forall j, k \in \mathbb{N}$. We assume that the initial state $x_{0}$ is a Gaussian random variable with zero mean and covariance $\Pi_{0}\geq 0$. Also $x_{0}$ is assumed to be uncorrelated with $\omega_{k}$ and $\nu_{k}$. We assume that $(A, C)$ is  observable and $(A,\sqrt{Q})$ is controllable.
	
	In cyber-physical systems, sensors are assumed to be intelligent \cite{Hovareshti2007} to implement some simple calculations. Thus, after making an observation at time $k$, the Kalman filter is used by the intelligent sensor to calculate the estimation of state $x_{k}$ locally.
	Denote $\hat{x}_{k}^{s}$ and $P^{s}_{k}$ as the local minimum mean-squared error (MMSE) estimate of the state $x_{k}$ and the corresponding error covariance:
	$
	\hat{x}_{k}^{s}=\mathbb{E}[x_{k}|y_{1},y_{2},...,y_{k}],
	P^{s}_{k}= \mathbb{E}[(x_{k}-\hat{x}_{k}^{s})(x_{k}-\hat{x}_{k}^{s})^{'}|y_{1},y_{2},...,y_{k}].
	$
	These terminologies are computed by the standard Kalman filter and the iteration begins with $\hat{x}_{0}^{s}=0$ and $P^{s}_{0}=\Pi_{0}$. For the sake of simplicity, the Lyapunov and Riccati operators $h$ and $\widetilde{g}$: $\mathbb{S}_{+}^{n}\rightarrow \mathbb{S}_{+}^{n}$  as $h(X)\triangleq AXA'+Q$ and $\widetilde{g}(X)\triangleq X-XC'[CXC'+R]^{-1}CX$ are defined.
	Due to the fact that $P^{s}_{k}$ converges exponentially fast to a unique fixed point $\bar{P}$ from any initial condition \cite{Sinopoli2004}, we assume that $P^{s}_{k}=\bar{P}, k \geq 1,$
	where $\bar{P}$ is the steady state error covariance given by the unique positive semi-definite solution of
	$\widetilde{g} \circ h(\bar{P})=\bar{P}$.
	\subsection{Communication over Fading Channel}
	In a typical wireless communication channel the transmitted  signal suffers attenuation and distortion resulted from interference, shadowing, process noise, and fading, and all these factors can frequently result in packet dropout \cite{Wu2020}.
	
	The wireless communication network is generally time-varying. In an industrial setting this time-variant characteristic may be resulted from moving machines, buildings,
	obstacle, and so forth,  the receptor or the sender
	built on a moving device can also result in the time-variable property of the channel. Hence, except for the
	propagation path loss, channels also suffer
	shadowing and small-scale fading.
	
\begin{figure}
		\centering
		\includegraphics[width=0.50\textwidth]{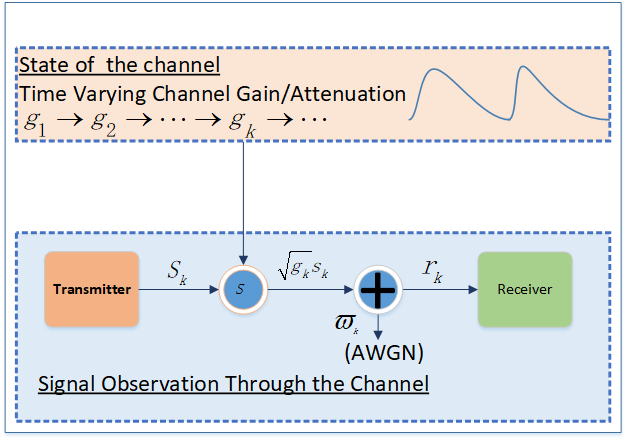}\\
		\caption{Fading Channel Model}\label{fig2}
	\end{figure}

	The architecture of wireless communication channel is depicted in Fig. \ref{fig2}. All frequency elements of the communication packet suffer a analogical fading or damping. As a result, the obtained signal at the receptor is provided as
	\begin{align}
	r_{t}=\sqrt{g_{t}}s_{t}+\varpi_{t},
	\end{align}
	where $s_{t}$ denotes the channel input which has an average energy constraint $ E[s^{2}_{t}]\leq P$.  $r_{t}$ represents the channel output;  ${\varpi_{t}}$ is an additive white Gaussian noise (AWGN) whose mean is zero and variance is $\sigma_{\varpi}^{2}$. Throughout the article, we will refer to the attenuation ${g_{t}}$ simply as the fading channel gain.
	
	Considering that the main characteristics of fading channels is a correlated, time-varying stochastic process. Specifically, the wireless transmission channel is a dynamic system, and the fading channel gain ${g_{t}}$ is a random process that transforms over time. An easy-to-handle mathematical model is needed to precisely describe the dynamic characteristics of time-varying fading channel.
In this paper, the common scheme such as finite state Markov chain (FSMC) is utilized to model wireless fading channels.  It is assumed throughout the paper that the following assumption holds:
	\begin{assumption}
		$g_{k}$ is the physical interpretation for channel quality and takes value in a finite set $\Xi$ whose elements are denoted by $\{ Z_{1},Z_{2},\cdots,Z_{l} \}$. Without loss of generality, we assume that $Z_{1}<Z_{2}<Z_{3}<\cdots<Z_{l}$.
		\begin{enumerate}
			\item $\Xi=\{Z_{k}\}_{k=1}^{l}$ is an ergodic Markov chain; We denote by $\mu$  the row vector of stationary probability distribution for  the Markov chain $\Xi$, and by $\mu_{i}>0$ its $i$-th entry corresponding to state $Z_{i}\in \Xi$.	
			\item The one-step transition probability for this chain is denoted by
			$
			\Pi(\cdot |\cdot): \Xi \times \Xi \longmapsto [0,1],
			$
			where $\Pi(\cdot |\cdot)$ is known a \textit{priori}. Also, $\Pi(\cdot |\cdot)$ is aperiodic and irreducible.	
			\item The channel is block fading, i.e., the channel gain $g_{k}$ keeps unchanged during each signal transmission but transforms from block to block.
			\item The accepter can verify the the received signal. Only the signal packets reconstructed error-free are identified as successful reception.
		\end{enumerate}
	\end{assumption}
	Finite state channel  models (FSCM) have been widely accepted as an effective method to describe the related structure of the fading channel \cite{Sadeghi2008}.  The FSCM is characterized by a deterministic
	or probabilistic function of a first-order Markov chain,
	where each element  may be related to a particular channel
	state.
	The Markov state process is in general stationary,  therefore,  the state transition probability is time-independent.  And the distribution of  initial state is assumed to be $u=[\mu_{1},\mu_{2},...,\mu_{l}]$. This is a common assumption in a real communication scenario for channel statistics are not fast-changing over time \cite{Sadeghi2008}.
	Random data packet dropout  will take place due to fading and interference. To describe this scenario,
	we adopt an  Additive White Gaussian Noise (AWGN) network, which chooses  Quadrature Amplitude Modulation (QAM), to describe the communication between the transmitter and the receiver.  After that, the symbol error rate (SER) is tightly connected with the signal to noise ratio (SNR) as
	$SER=2Q(\sqrt{\alpha SNR}),$
	where $\alpha>0$ is a parameter and $Q(x)\triangleq \frac{1}{\sqrt{2\pi}}\int_{x}^{\infty}exp(-\eta^{2}/2)d\eta$.
	Consider a DoS attack congests the communication channel of the network.
	The corresponding SNR of the channel is transformed to SINR:
	$$
	SINR=\frac{\Psi^{s}g^{s}}{\Psi^{a}g^{a}+\sigma^{2}},
	$$
	where $\Psi^{s}$ is the transmitted energy used by the sensor, $g^{s}$ represents the channel gain for sensors, $\Psi^{a}$ is the interfering energy  from the attacker, $g^{a}$ represents the channel gain for attackers, and $\sigma^{2}$ represents  the additive white Gaussian noise.
	Thus taking account of both fading and additive white Gaussian noise in the transmission, whether the signals  transmitted by the sensor are successfully accepted by the remote center can be formulated as a binary random process ${\gamma_{k}},k\in \mathbb{N}$ (which indeed follows a Bernoulli distribution), in which $\gamma_{k}=0$ indicates that the packet has been lost.
	\begin{align}\label{gamma}
	&\gamma_{k} = \left\{ \begin{array}{ll}
	1, & \textrm{if data packet is received error-free},\\
	0 ,& \textrm{otherwise.
	}
	\end{array}\right.
	\end{align}
	Denote the energy levels of the sensor and the attacker, respectively, as:
	\begin{align*}
	\mathbb{P}_{\mathcal{S}} \triangleq [\mathcal{P}_{s_{1}},\cdots, \mathcal{P}_{s_{max}}] \quad \textrm{and} \quad
	\mathbb{P}_{\mathcal{A}} \triangleq [\mathcal{P}_{a_{1}},\cdots, \mathcal{P}_{a_{max}}].
	\end{align*}
	The successful packet reception is not only statistically determined by the power levels  but also by the channel gains of the sensor and the attacker.
	
	This finite state fading channel model generates an error symbol based on the following probabilistic mechanism. At time $k$, the chain state of two player is $g^{s}_{k}$, $g^{a}_{k}$, respectively. It generates an output symbol $\gamma_{k} \in \{0,1\}$ with probability
	\begin{equation}\label{dropout-rate}
	q(\mathcal{P}_{s_{k}},g^{s}_{k},\mathcal{P}_{a_{k}},g^{a}_{k})\triangleq Pr(\gamma_{k}=1|\mathcal{P}_{s_{k}},g^{s}_{k},\mathcal{P}_{a_{k}},g^{a}_{k}),
	\end{equation}
	where $\mathcal{P}_{s_{k}}\in \mathbb{P}_{\mathcal{S}}$,  $\mathcal{P}_{a_{k}}\in \mathbb{P}_{\mathcal{A}}$, $g^{s}_{k}\in \Xi $ and $g^{a}_{k}\in \Xi $. Based on the state process and channel gains, the error symbol is memoryless, this is
	\begin{align*}
	Pr(\gamma_{1:k}|\mathcal{P}_{s_{1:k}},\mathcal{P}_{a_{1:k}},g^{s}_{1:k},g^{a}_{1:k})=\prod_{i=1}^{k}Pr(\gamma_{i}|\mathcal{P}_{s_{i}},\mathcal{P}_{a_{i}},g^{s}_{i},g^{a}_{i}).
	\end{align*}
	To avoid trivial problems, we assume that the following equation holds:
	$
	\min_{\mathcal{P}_{s_{k}},g^{s}_{k},\mathcal{P}_{a_{k}},g^{a}_{k}} q(\mathcal{P}_{s_{k}},g^{s}_{k},\mathcal{P}_{a_{k}},g^{a}_{k})>1-\frac{1}{\rho(A)^{2}},
	$
	where $A$ is the system matrix and $\rho(A)$ represents the spectral radius of $A$. Under the above sufficient condition, the expected estimation error covariance is bounded.
	While the above condition is not satisfied, the attacker would jam the channel with a certain interfering energy continuously to obtain an unbounded estimation error, that is to say, the attacker dominates the estimation process and thereby there exists no equilibrium for the attacker and sensor.
	
	\subsection{Remote State Estimation}
	Denote  $\hat{x}_{k}$ and $P_{k}$ as state estimate and corresponding error covariance of the system in the remote center side. According to the work \cite{Sinopoli2004}, they can be obtained via the following procedure:
	\begin{align}
	&\hat{x}_{k} = \left\{ \begin{array}{ll}
	\hat{x}^{s}_{k}, & \textrm{$\gamma_{k}=1$} ,\\
	A\hat{x}_{k-1},& \textrm{$\gamma_{k}=0$}.
	\end{array}\right.
	\end{align}
	And $P_{k}$ is computed as follows:
	\begin{align}\label{error}
	&P_{k} = \left\{ \begin{array}{ll}
	\bar{P}, & \textrm{$\gamma_{k}=1$} ,\\
	h(P_{k-1}),& \textrm{$\gamma_{k}=0$},
	\end{array}\right.
	\end{align}
	where $\bar{P}$ represents the steady-state error covariance.
	Without loss of generality, we assume that the initial packet $\hat{x}^{s}_{0}$ is acquired by the remote center and hence $P_{0}=\bar{P}$.

	In order to express succinctly, we define a random variable $\tau_{k}\in \mathbb{N}$ as the  duration between two  transmission:
	\begin{align}\label{tau}
	\tau_{k}\triangleq k-\max_{0 \leq t \leq k}\{t: \gamma_{t}=1 \},
	\end{align}
	which represents the distances between the time $k$ and the most recent time that the transmitted signal is acquired by the remote center. The estimation error covariance is associated with the holding time and the relationship is described as
	$
	P_{k}=h^{\tau_{k}}(\bar{P}),
	$
	and the holding time updates as follows:
	\begin{align}
	&\tau_{k} = \left\{ \begin{array}{ll}
	0, & \textrm{$\gamma_{k}=1$} ,\\
	\tau_{k-1}+1,& \textrm{$\gamma_{k}=0$}.
	\end{array}\right.
	\end{align}
	Note that the energy and channel gains of the two agents (the sensor and the attacker), i.e., $\mathcal{P}_{s_{k}},\mathcal{P}_{a_{k}},g^{s}_{k},g^{a}_{k}$ are given, and the sequence of stochastic variable $\tau_{k}$ forms a Markov chain with the transition probability matrix $\mathbb{T}$ given as follows:
	\begin{gather}\label{transition matrix}
	\mathbb{T}=
	\begin{bmatrix}
	q & 1-q & 0 & 0 & 0 & \cdots  \\
	q & 0 & 1-q & 0 & 0 & \cdots \\
	q & 0 &0 & 1-q & 0 & \cdots\\
	\vdots & \vdots &\vdots &\vdots &\ddots& \cdots  \\
	\end{bmatrix}
	\end{gather}
	where the elements of  $\mathbb{T}$ represent the transition probability from the state $\tau_{k}=i$ to $\tau_{k+1}=j$, and the missing elements are $0$. And the probability $q=q(\mathcal{P}_{s_{k}},g^{s}_{k},\mathcal{P}_{a_{k}},g^{a}_{k})$ is given by \eqref{dropout-rate}.
	In the following subsection, we formulate the problem of interest. \vspace{-3mm}
	\subsection{Problem of Interest}
	Depending on the channel knowledge of  sensors and  attackers, we investigate how to adopt game theory to develop a transmission schedule for sensors and attackers under Nash equilibrium, where the choice of each player is the best response to the choice of other opponents.  The sensor and the attacker are assumed to be rational and they will always adopt the behavior that offers the highest expected reward. Therefore, in this work, our goal is to find the optimal strategy for both sides such that no one obtains more through unilateral deviation. The complete information for the stochastic game is constructed as follows.\vspace{-3mm}
	\section{Main Results}\label{sec:stochastic-game}
	\subsection{Stochastic Game Description}
	In this subsection, we first model the schedule of scheduling energy-efficient actions in an infinite time horizon. In other words, we investigate how to arrange the jamming power (or transmission energy) for the attacker (or the sensor) in a stochastic game framework.
	In this stochastic game framework, players decide action simultaneously. We assume that the state set and action set are discrete. The formal definition of this stochastic game is provided in the following:
	\begin{definition}\textit{
			An attacker-sensor stochastic game $\Gamma$ consists of a tuple $<\mathcal{S},\mathcal{A}^{1},\mathcal{A}^{2},r^{1},r^{2}, p>$, where $\mathcal{S}$ represents the state set, $\mathcal{A}^{i}$  stands for  the action set for player $i$.    $r^{i}:\mathcal{S}\times \mathcal{A}^{1} \times \mathcal{A}^{2}\rightarrow R$ represents the reward function for player $i$ ($i=1,2$), $p:\mathcal{S}\times \mathcal{A}^{1} \times \mathcal{A}^{2}\rightarrow \Delta(\mathcal{S})$ stands for the transition probability mapping, where $\Delta(\mathcal{S})$ represents the set of probability distribution over state set $\mathcal{S}$.}
	\end{definition}
	
	In our problem, we provide the following specific content to describe the sensor-attacker game.
	
	\textbf{\textit{Player}}: We assume that the attacker and the sensor are all rational players. Denote by $\mathcal{I}^{1}$ the attacker and $\mathcal{I}^{2}$ the sensor. The attacker and the sensor decide  the best action among all available choice for them in terms of their own goals.
	
	\textbf{\textit{State}}: The state space $\mathcal{S}$ can be expressed as $\mathcal{S}=\{(\tau_{k},g_{s}, g_{a})\}$, where $\tau_{k} \in \mathbb{N}$, $g_{s}$ and $g_{a}$ are the channel gain of sensor and attacker, respectively.  
	Notice that the holding time is tightly associated with the estimation error covariance $P_{k}$ as $P_{k}=h^{\tau_{k}}(\bar{P})$.
	
	\textbf{\textit{Action}}: Denote by $\mathcal{A}^{i}$  the action (or pure strategy) space for player $i$, $i=1,2$.  At time $k$, the attacker decides the interfering energy $a_{k}\in \mathbb{P}_{\mathcal{A}}\triangleq [\mathcal{P}_{a_{1}},\cdots, \mathcal{P}_{a_{max}}]$, and the sensor selects the transmission energy  $b_{k}\in\mathbb{P}_{\mathcal{S}}\triangleq [\mathcal{P}_{s_{1}},\cdots, \mathcal{P}_{s_{max}}].$ Denote by  $(a_{k},b_{k})$ the joint action (or pure strategy) at time $k$.
	
	\textbf{\textit{Transition probability}}:
	Note that the holding time $\tau_{k}$ forms a Markov chain in which the transition matrix is presented in \eqref{transition matrix}.
	And the transition probability is determined by  the packet arrival probability in \eqref{dropout-rate}. From (15), $s_{k}$ has the Markov property. 
	Denoted by $s_{k+1}=(\tau_{k+1}, g_{s}, g_{a}), s_{k}=(\tau_{k}, g'_{s}, g'_{a})\in \mathcal{S}$ and $a\in \mathbb{P}_{\mathcal{A}},b\in \mathbb{P}_{\mathcal{S}}$, $g_{s}=t$, $g'_{s}=t'$, $g_{a}=e$, $g'_{a}=e'$ where $t, e, t', e' \in \Xi$, then $\forall k \geq 0$, $Pr(s_{k+1}|s_{k}, a_{k}=a, b_{k}=b) =$
	\begin{eqnarray*}
		\begin{aligned}
			&\left\{ \begin{array}{ll}
				[1-q(a,t',b,e')]u(t')u(e'), & \textrm{$s_{k+1}=(\tau_{k}+1,t,e)$}, \\
				q(a,t',b,e')u(t')u(e'),& \textrm{$s_{k+1}=(0, t, e)$},\\
				0, &\textrm{otherwise},
			\end{array}\right.
		\end{aligned}
	\end{eqnarray*}
	where $u(g_{s}),u(g_{a}) \in [0,1]$ stands for  the stationary distribution probability  of channel gain $g_{s}$ and $g_{a}$, respectively.
	
	\textbf{\textit{Reward function:}}
	Denote $r^{i}$ the immediate payoff function of player $i$ with $r^{i}:\mathbb{N} \times \mathcal{A}^{1} \times \mathcal{A}^{2}\rightarrow R$.
	The payoff function of attacker is provided as
	$
	r^{1}(m,a,b)=Tr[h^{m}(\bar{P})]+\alpha_{s}b-\alpha_{a}a.
	$ And the sensor's  immediate reward function is denoted as $r^{2}(m,a,b)=-r^{1}(m,a,b)$.
	
	
	Defining  the infinite time horizon discounted sum of rewards as follows:
	\begin{equation}
	\begin{aligned}\label{JAJS}
	v^{1}(s,\pi^{1},\pi^{2})=\sum_{k=0}^{+\infty}\beta^{k} E[r^{1}|\pi^{1},\pi^{2},s_{0}=s],
	\end{aligned}
	\end{equation}
	\begin{equation}
	\begin{aligned}
	v^{2}(s,\pi^{1},\pi^{2})=\sum_{k=0}^{+\infty}\beta^{k} E[r^{2}|\pi^{1},\pi^{2},s_{0}=s],
	\end{aligned}
	\end{equation}
	where $v^{1}(s,\pi^{1},\pi^{2})$ and $v^{2}(s,\pi^{1},\pi^{2})$ represent the value for state $s$ under the circumstance when the attacker adopts the strategy $\pi^{1}$ and the sensor adopts strategy $\pi^{2}$. A \textit{strategy} is a schedule  for the players to take action. Here
	\begin{align}\label{strategypi}
	\pi^{i}=(\pi^{i}_{0},...,\pi^{i}_{k},...)
	\end{align} for $i=1,2$ is defined on the whole process, where $\pi^{i}_{k}$  stands for  the \textit{decision schedule}.  A decision schedule  is a mapping $\pi^{i}_{k}:\boldsymbol{H}_{k}\longmapsto\Delta(A)$, where $\boldsymbol{H}_{k}$ represents  the space of the history of the probability  before  time $k$, with each $H(k)\in \boldsymbol{H}_{k}, H_{k}=(s_{0},a^{1}_{0},a^{2}_{0},...s_{k-1},a^{1}_{k-1},a^{2}_{k-1},s_{k})$, and $\Delta(A)$ stands for  the set of probability distributions over the player's actions.
	
	The process of game is summarized as follows. When the process is in state $s_{k}$ at time $k>0$, they independently and simultaneously take action $(a_{k},b_{k})$ from the available action set on the basis of stochastic stationary policy that will be explained in the following subsection.  Therefore,  the attacker obtains an immediate reward $r^{1}(s_{k},a_{k},b_{k})$ (or cost $-r^{1}(s_{k},a_{k},b_{k})$ is generated for the sensor simultaneously); the process jumps to a new state $s_{k+1}$ with a conditional probability  depended on $Pr(s_{k+1}|s_{k},a_{k},b_{k})$. The objective of the attacker (or the sensor) is to maximize its rewards in regard to the discounted sum standard  $v^{1}(s,\pi^{1},\pi^{2})$, which is defined in \eqref{JAJS}.
	
	Therefore, the attacker-sensor  game is constructed to formulate the conflicting characteristics between the attacker and the sensor \cite{Filar-Vrieze}.  Besides, the specific description of  game elements are provided. We present in the following subsection  that the Nash equilibrium of attacker-sensor game is
	existed.
	\vspace{-2mm}
	\subsection{Existence of Equilibrium Strategies}
The strategy $\pi^{i}$ is referred to  a \textbf{\textit{stationary strategy}} if $\pi^{i}_{k}=\bar{\pi}$, i.e., the decision schedule is time-independent  and determined only by  the current state $s\in \mathcal{S} $. $\pi^{i}$ is called a \textbf{\textit{behavior strategy}} if its decision schedule  is determined by the previous information of game, $\pi^{i}_{k}=f(H_{k})$.
	
	The Nash equilibrium is composed of  a joint strategy where the attacker (the sensor) has  a best response to the sensor (the attacker). For this attacker-sensor game, the attacker's (or the sensor's) schedule is defined in the infinite time horizon.
	\begin{definition}
		\textit{In our attacker-sensor  game $\Gamma$, a Nash equilibrium point is composed  of two strategies ($\pi^{1}_{*},\pi^{2}_{*}$) such that for all $s \in \mathcal{S}$ we have
			\begin{align*}
			v^{i}(s,\pi^{1}_{*},\pi^{2}_{*})\geq v^{i}(s,\pi^{1},\pi^{2}_{*}),
			v^{i}(s,\pi^{1}_{*},\pi^{2}_{*})\geq v^{i}(s,\pi^{1}_{*},\pi^{2})
			\end{align*}
			for all $\pi^{i} \in \Omega^{i}$ and $i=1,2$, where $\Omega^{i}$ represents the set of strategies for player $i$.
	}\end{definition}
	
	However, the strategy pair which forms a Nash equilibrium point may be a stationary strategy or a behavior strategy. We prove that  an Nash equilibrium in stationary strategies is always existed as follows.
	\begin{lemma}[cf.\cite{Fink1964}]\textit{
		Under the stationary strategy, every n-player discounted game has at least one Nash equilibrium point.}
	\end{lemma}
	
	
	Throughout  this work, we focus on stationary strategy. Non-stationary strategy, i.e., behavior strategy, which is based on the action history, are rather complicated, and relatively less research in the game framework.
	Though the existence of a stationary Nash equilibrium has been demonstrated, it is still hard to establish practically a look-up table about the optimal strategy. Thus, we provide the calculation method in the following.
	\subsection{Practical Implementation}
	In this subsection, we provide a modified Q-learning algorithm, which is called Nash Q-learning algorithm, to find the optimal solution  $(\pi^{1}_{*},\pi^{2}_{*})$  for sensors and attackers. 
	To adopt Q-learning to multi-players, we should recognize that it is necessary to investigate joint actions instead of  only individual  action. As to the attacker-sensor game, the Q-function of each player is $Q(s,a_{1},a_{2})$, instead of the
	single player Q-function $Q(s,a)$.  Based on the extended form of Q-function and concept of Nash equilibrium,  the Nash Q-value is defined as the expected sum of  discounted reward when all players adopt the Nash equilibrium strategy from the next state of arrival. The above definition is different from the single player scenario in which the future rewards are depended only on the player's personal optimal scheme.  To be more precise, we define
	$Q^{i}_{*}$ as a Nash Q-function for player $i$.
	
	\begin{definition}
		\textit{The Nash Q-function of player $i$ is defined over $(s,a_{1},a_{2})$, as the sum of player's current payoff plus its future rewards when all players adopt a joint Nash equilibrium strategy.  More specifically,
		\begin{align*}
			Q^{i}_{*}(&s,a_{1},a_{2})
			\\&=r^{i}(s,a_{1},a_{2})
			+\beta\sum_{s^{'}\in \mathcal{S}}Pr(s^{'}|s,a_{1},a_{2})v^{i}(s^{'},\pi_{*}^{1},\pi_{*}^{2}),
\end{align*}
			where $(\pi^{1}_{*},\pi^{2}_{*})$ constitutes the joint Nash equilibrium strategy, $r^{i}(s,a_{1},a_{2})$ represents player $i$'s immediate payoff in state $s$ and with the action pair $(a_{1},a_{2})$, $v^{i}(s^{'},\pi_{*}^{1},\pi_{*}^{2})$ stands for  player $i$'s entire discounted reward over infinite time horizon beginning with state $s^{'}$ provided that the player adopts  the Nash equilibrium strategy.}
	\end{definition}

	In Nash Q-learning algorithm, the player makes an attempt to acquire its equilibrium Q-values, beginning with any initial state. For this purpose,  the player keeps a record of opponent players' Q-values and takes advantages of  that message to calculate its own Q-values. The updating manner is depended on the fact that opponents adopt their equilibrium strategy in every state.
	Before designing the algorithm,  we need the following definition.
	\textit{\begin{definition}
			We define a two-player stage game as $(U^{1}, U^{2})$, where $U^{k}$ represents player $k$'s immediate reward function over the entire joint pairs for $k=1,2$. More specifically,  $U^{k}=\{r^{k}(s,a_{1},a_{2})|  a_{1}\in \mathcal{A}^{1}, a_{2}\in \mathcal{A}^{2}\}$, and $r^{k}$ stands for the payoff of player  $k$.
	\end{definition}}
	
	Denote  by $\pi^{-k}$ the product of strategies of all players except for $k$, $\pi^{-k}\triangleq \pi^{1}\cdots \pi^{k-1}\cdot \pi^{k+1}\cdots \pi^{n}$.
	
	\begin{definition}
		\textit{A joint strategy $(\pi^{1}, \pi^{2})$ constitutes a Nash equilibrium for the stage game $(U^{1},...,U^{n})$ if,  for $k=1,...,n$,
			$
			\pi^{k}\pi^{-k}U^{k}\geq \vartheta^{k}\pi^{-k}U^{k}
			$for all $\pi^{k}\in \vartheta^{k}(\mathcal{A}^{k})$ and $\vartheta^{k}$ is an arbitrary strategy of player $k$.
		}
	\end{definition}

	The baseline method for attacker-sensor game is to adopt the Nash equilibrium. In a Nash equilibrium point, each participant actually keeps a right expectation about the opponent players' responses, and behaves reasonably  according to  this expectation. 
	
	The goal of this paper  is to obtain a mixed strategy pair $\{\pi^{1}_{*},\pi^{2}_{*}\}$ at a Nash equilibrium, where $\pi^{1}_{*}=[\pi^{*}_{\mathcal{A}0},\cdots,\pi^{*}_{\mathcal{A}max}]$ and $\pi^{2}_{*}=[\pi^{*}_{\mathcal{S}0},\cdots,\pi^{*}_{\mathcal{S}max}]$.
	The optimal Q-value function for the attacker is defined as
	\begin{eqnarray*}
		\begin{aligned}
			Q^{1}_{*}(s,a_{1},a_{2})= r^{1}(s,a_{1},a_{2})+\beta \max_{\pi^{1}} \min_{\pi^{2}} &\sum_{a'_{1} }\sum_{a'_{2} }Q^{1}_{*}(s',a'_{1},\\&  a'_{2}) \pi^{1}(a^{'}_{1})\pi^{2}(a^{'}_{2}),
		\end{aligned}
	\end{eqnarray*}
	where $\pi^{1}(a^{'}_{1})$ and $\pi^{2}(a^{'}_{2})$ are the probabilities of choosing  $a^{'}_{1}$ in strategy $\pi^{1}$ for the attacker and adopting $a^{'}_{2}$ in strategy $\pi^{2}$ for the sensor, respectively. $Q^{1}_{*}(s,a_{1},a_{2})$ is thought as the expected payoff for attackers taking action $a_{1}$ and sensors executing $a_{2}$, and then they execute the optimal policy. While the optimal Q-value is acquired, we can easily find the optimal policies $\pi^{1}_{*}$.
	\IncMargin{1em}
	\begin{algorithm}	
		\caption{Nash Q-learning algorithm}
		\SetAlgoLined	
		\SetKwInOut{Input}{Input}
		\SetKwInOut{Output}{Output}
		\begin{flushleft}
			\textbf{Input:}{ Finite state channel gain set $\Xi$, action space for the attacker $\mathcal{P}_{\mathcal{A}}$  and the sensor $\mathcal{P}_{\mathcal{S}}$, packet arrival function $q(\cdot)$, learning rate $\alpha_{k}$, discount factor $\psi$.\;
		}\end{flushleft}
		\begin{flushleft}\textbf{Output:}
			Nash Q-function value $Q^{i}_{*}(\cdot,\cdot,\cdot)$, Nash equilibrium strategy $(\pi^{1}_{*},\pi^{2}_{*})$.
		\end{flushleft}
		\begin{flushleft}\textbf{Initialize:}
			Let time $k=0$, define the initial state $s_{0}$;  Assign the learning player be indexed $i$; For all $s\in \mathcal{S}$
			and $a_{i}\in \mathcal{A}^{i}$, $i=1,2$, let $Q^{i}_{0}(s,a_{1},a_{2})=0$.
		\end{flushleft}
		\Repeat
		{\text{convergence}}
		{1: Choose action $a^{k}_{i}$ at time $k$.\; \\
			2: Observe the reward function of all player $r^{1}_{k}$, $r^{2}_{k}$, action $a^{k}_{1}$, $a^{k}_{2}$, and $s_{k+1}=s'$\;\\	
			3: \For {each player $j=1,2$}{
				\begin{eqnarray*}
					\begin{aligned}
						&Q^{i}_{k+1}(s)=(1-\alpha_{k})Q^{i}_{k}(s)+\alpha_{k}[r^{i}_{k}+\beta \textit{Nash} Q^{i}_{k}(s')]
					\end{aligned}
				\end{eqnarray*}
				where $\alpha_{k}\in(0,1)$ represents the learning rate, and $\textit{Nash} Q^{j}_{k}$ is defined in \eqref{nash}. \;\\
			}
			4:	Decay the learning rate $\alpha_{k}$ and update the time $k=k+1$.
		}
	\end{algorithm}
	\DecMargin{1em}
	
	Similarly, the Q-value for sensors is exactly the opposite in the zero-sum game, 
	%
	i.e.,  $Q^{1}_{*}(s,a_{1},a_{2})=-Q^{2}_{*}(s,a_{1},a_{2})$.
	More specifically,  the participants will provide the initial values of $Q(s,a_{1},a_{2})$ for all  $a_{1}\in \mathbb{P}_{\mathcal{A}}$,  $a_{2}\in \mathbb{P}_{\mathcal{S}}$, and $s\in \mathcal{S}$. Without loss of generality, let $Q^{i}_{0}(s,a_{1},a_{2})=0$. At each step, player $i$ acquires the current state, and chooses its action.  Then, it collects its own payoff,  all opponent players' action, opponents' rewards, and the new state $s'$. When an action is taken, $r^{1}$ and $r^{2}$ are obtained. The corresponding Q-values of attacker are calculated as follows:
	\begin{eqnarray*}
		\begin{aligned}
			Q^{1}_{k+1}(s,a_{1},a_{2})=&(1-\alpha_{k})Q^{1}_{k}(s,a_{1},a_{2})+\alpha_{k}\Big[r^{i}_{k}+\beta \max_{\pi^{1}}\\& \min_{\pi^{2}}
			\sum_{a_{1} }\sum_{a_{2} }Q^{1}_{k}(s',a'_{1},a'_{2}) \pi^{1}(a'_{1})\pi^{2}(a'_{2})\Big],
		\end{aligned}
	\end{eqnarray*}
	where $\alpha_{k}\in[0,1)$ represents the learning rate and it decays over time.
	Notice that the attacker updating its Q-value can be represented  in a general form:
	\begin{eqnarray}\label{updaterule}
	\begin{aligned}
	Q^{1}_{k+1}(s,a)=(1-\alpha_{k})Q^{1}_{k}(s,a)+\alpha_{k}[r^{i}_{k}+\beta \textit{Nash} Q^{1}_{k}(s')]
	\end{aligned}
	\end{eqnarray}
	where $a=(a_{1},a_{2})$ and
	\begin{align}\label{nash} \textit{Nash}Q^{1}_{k}(s')=\max_{\pi^{1}} \min_{\pi^{2}} \pi^{1}(s')\pi^{2}(s')Q^{1}_{k}(s').
	\end{align}
	The corresponding Q-values of the sensor $Q^{2}_{*}(s,a_{1},a_{2})$ can be obtained in a similar way.
	Distinct means of choosing from all of the equilibrium points will generally cause different updates.
	\textit{Nash}$Q^{i}_{k}(s')$ stands for player $i$'s reward in state $s'$ for the chosen equilibrium. Notice that $\pi^{1}(s')\pi^{2}(s')Q^{i}_{k}(s')$ is a scalar.
	In our implementation, we find the  Nash equilibria  by using the Lemke-Howson method \cite{Cottle1992}, which is rather efficient in reality in spite of exponential worst-case behavior. The Lemke-Howson method is based upon a simple pivoting strategy, which corresponding to following a path whose endpoints is a Nash equilibrium.

	Player $i$ need to know $Q^{1}_{k}(s')$ and $Q^{2}_{k}(s')$ so as to compute the Nash equilibrium $(\pi^{1}(s'),\pi^{2}(s'))$. Note that information about other players' Q-values is not known. Player $i$ makes guesses about those Q-functions at the initial moment of game, for instance, $Q^{i}_{0}(s,a_{1},a_{2})=0$ for all $i$, $s,a_{1},a_{2}$. With the proceeding of game, player $i$ collects other players' immediate rewards and previous actions. Those information is utilized to update player $i$'s guesses on opponent players' Q-functions. Player $i$ updates its conjectures about player $j$'s Q-function based on the same updating rule \eqref{updaterule} it applied to itself,
	\begin{eqnarray}\label{rulej}
	\begin{aligned}
	Q^{j}_{k+1}(s,a)=(1-\alpha_{k})Q^{j}_{k}(s,a)+\alpha_{k}[r^{j}_{k}+\beta \textit{Nash} Q^{j}_{k}(s')].
	\end{aligned}
	\end{eqnarray}
	Note that $\alpha_{k}=0$ for $(s,a_{1},a_{2})\neq(s_{k},a_{1}^{k},a_{2}^{k})$.
	It only updates the entry corresponding to the same state and actions selected by players. Such updating is referred to asynchronous updating. This learning algorithm is summarized in \textbf{\textit{Algorithm 1}}.\vspace{-2mm}
	\subsection{Convergence analysis}
	We now prove $Q^{i}_{k}$ of the player $i$ converges  to the optimal Q-value $Q^{i}_{*}$. The value of $Q^{i}_{*}$ is decided on the common strategy of all participants. This means that we need to demonstrate $(Q^{1}_{k},Q^{2}_{k})$ converges to $(Q^{1}_{*},Q^{2}_{*})$.
	
	The convergence analysis of our algorithm is based on the following three basic assumptions which is connected with infinite sampling and decaying of learning rate \cite{Hu1998}:
	
	\begin{assumption}\textit{
			The learning rate $\alpha_{k}$ satisfies $\sum_{k=0}^{\infty}\alpha_{k}=+\infty$ and $\sum_{k=0}^{\infty}\alpha_{k}^{2}<+\infty$, where $0 \leq\alpha_{k}\leq 1$.}
	\end{assumption}
	
	\begin{assumption}\textit{ All states $s\in \mathcal{S}$ and actions $a^{i}\in\mathcal{A}^{i}$ for $i=1,2$ have been visited infinitely often.}
	\end{assumption}
	\begin{assumption}\label{covcon}
		\textit{
			One of the following conditions holds during learning.
			\begin{condition}
				Every stage game $(Q^{1}_{k}(s),Q^{2}_{k}(s))$, for all $k$ and $s$, possesses a global optimal point, and players' rewards in this equilibrium are utilized to calculated their Q-functions.
			\end{condition}
			\begin{condition}
				Every stage game $(Q^{1}_{k}(s),Q^{2}_{k}(s))$, for all $k$ and $s$, possesses a saddle point, and players' rewards in this equilibrium are utilized to calculated their Q-functions.
		\end{condition}}
	\end{assumption}
	
	\textit{Assumption $2$} implies the decaying of the learning rate. In order to guarantee \textit{Assumption $2$} satisfied,  the learning rate is scheduled to be a non-zero decreasing function and the current state-action pair.  \textit{Assumption $3$} is satisfied with a great deal of update processes since the  action set is finite.
	
	Now, we prove that the process incurred by NashQ updates in
	\eqref{updaterule} converges to Nash Q-values in the following theorem.
	\begin{theorem}
		When the Assumption $2-4$ are satisfied, the sequence $Q_{k}=(Q^{1}_{k},Q^{2}_{k})$, updated by
		$Q^{i}_{k+1}(s,a_{1},a_{2})=(1-\alpha_{k})Q^{i}_{k}(s,a_{1},a_{2})+\alpha_{k}(r^{i}_{k} +\beta\psi^{1}(s')\psi^{2}(s')Q^{i}_{k}(s')),$  where $i=1,2,$ and $(\psi^{1}(s'),\psi^{2}(s'))$ is the calculated Nash equilibrium solution for the stage game $(Q^{1}_{k}(s'),Q^{2}_{k}(s'))$, converges to the Nash Q-value $Q_{*}=(Q^{1}_{*},Q^{2}_{*})$.
	\end{theorem}
	\begin{IEEEproof}
		See appendix A.
	\end{IEEEproof}
	\begin{remark}
		The convergence of Nash Q-learning algorithm for zero-sum stochastic game is guaranteed if either \textit{Condition} 1 or \textit{Condition} 2 satisfies.
		However, such conditions are not necessary \cite{Vamvoudakis2015}.
		With respect to the practical example, experiments with a large number of two-player games suggested that such limitations  in the game framework are not necessarily required and results all show the empirical convergence of the Q-value.
	\end{remark}
	\newcounter{mytempeqncnt}
	\begin{figure*}[!ht]
		\normalsize
		\setcounter{mytempeqncnt}{\value{equation}}
		\renewcommand{\theequation}{A\arabic{equation}}
		\setcounter{equation}{0}
		\begin{equation}\label{Q1}
		Q_{*}^{i}(s',\overrightarrow{a_{1}})
		=r^{i}(m+1,a_{1}^{+},a_{2}^{+})+\beta\big[ q(a_{1}^{+},a_{2}^{+}, g'_{s},g'_{a})u(g'_{s})u(g'_{a})v_{*}^{i}(0)
		+(1-q(a_{1}^{+},a_{2}^{+}, g'_{s},g'_{a}))u(g'_{s})u(g'_{a})v_{*}^{i}(m+2)\big]
		\end{equation}
		\begin{equation}\label{Q3}
		Q_{*}^{i}(s',\overrightarrow{a_{2}})=r^{i}(m+1,a_{1}^{-},a_{2}^{-})+\beta\big[ q(a_{1}^{-},a_{2}^{-},g'_{s},g'_{a})u(g'_{s})u(g'_{a})v_{*}^{i}(0)
		+(1-q(a_{1}^{-},a_{2}^{-},g'_{s},g'_{a}))u(g'_{s})u(g'_{a})v_{*}^{i}(m+2)\big]
		\end{equation}
		\begin{equation}\label{Q2}
		Q_{*}^{i}(s,\overrightarrow{a_{2}})=r^{i}(m,a_{1}^{-},a_{2}^{-})+\beta\big[ q(a_{1}^{-},a_{2}^{-}, g_{s}, g_{a})u(g_{s})u(g_{a})v_{*}^{i}(0)
		+(1-q(a_{1}^{-},a_{2}^{-}, g_{s}, g_{a}))u(g_{s})u(g_{a})v_{*}^{i}(m+1)\big]
		\end{equation}
		\begin{equation}\label{Q4}
		Q_{*}^{i}(s,\overrightarrow{a_{1}})=r^{i}(m,a_{1}^{+},a_{2}^{+})+\beta\big[ q(a_{1}^{+},a_{2}^{+}, g_{s}, g_{a})u(g_{s})u(g_{a})v_{*}^{i}(0)
		+(1-q(a_{1}^{+},a_{2}^{+}, g_{s}, g_{a}))u(g_{s})u(g_{a})v_{*}^{i}(m+1)\big]
		\end{equation}
		\setcounter{equation}{\value{mytempeqncnt}}
		\hrulefill
	\end{figure*}
	\setcounter{equation}{33}

	\subsection{Strictly increasing  Structure of Optimal Nash Stationary Strategies }
	In this subsection, we aim at establishing the optimal \textit{Nash} stationary strategies with special structure. The significance of conclusion in regard to the optimality of structured strategies consists in their attraction to deciders, their convenience in implementation, and their enabling efficient computation.  For example, if we have proved that the optimal  \textit{Nash} stationary  strategies are strictly increasing functions of state, which means that the higher the current states, the larger  the  optimal action in the subsequent period. The special structure of the optimal \textit{Nash} stationary strategy is presented as follows.
	
	In the discounted attacker-sensor stochastic game, each player aims at maximizing the the sum of discounted  rewards. Player $i$ aims at maximizing
	\begin{align}
	v^{i}(s,\pi^{1},\pi^{2})=\sum_{k=0}^{+\infty}\beta^{k} E(r^{i}|\pi^{1},\pi^{2},s_{0}=s).
	\end{align}
	Note that we have defined the \textit{Equilibrium} strategy in \textit{Definition 3}.
	Based on the result in \textit{Lemma 1}, we obtain that our sensor-attacker game owns at least one Nash equilibrium point in stationary strategies. Besides, we employ the Nash Q-learning algorithm to acquire this optimal strategy at the Nash equilibrium.
	
	As to the attacker, the solution to the problem above is obtained by seeking  a  fixed point of the equation
	$v^{1}(s)=\max_{a_{1}}\min_{a_{2}}\{r^{1}(s,a_{1},a_{2})+\beta^{k}\sum_{s'\in \mathcal{S}}Pr(s'|s,a_{1},a_{2})
	v^{1}(s')\}.$
	Note that in \textit{Definition 4}, we have defined the Nash Q-function as
	\begin{align*}
	Q^{i}_{*}&(s,a_{1},a_{2})\\
	&=r^{i}(s,a_{1},a_{2})
	+\beta\sum_{s^{'}\in \mathcal{S}}Pr(s'|s,a_{1},a_{2})v_{*}^{i}(s',\pi_{*}^{1},\pi_{*}^{2}).
\end{align*}
	\begin{remark}
		Note that $Q^{i}_{*}(s,a_{1},a_{2})$ can be thought as the expected payoff for the player $i$ in which attackers take action $a_{1}$ and sensors execute action $a_{2}$, and then they follow the optimal schedule  thereafter.  If the optimal Q-value $Q^{i}_{*}(s,a_{1},a_{2})$ is acquired, the optimal policies $(\pi^{1}_{*}, \pi^{2}_{*})$ can be easily found. Therefore, the strictly increasing structure of optimal policies are analyzed based on the strict supermodularity of $Q^{i}_{*}(s,a_{1},a_{2})$ in the following.
	\end{remark}
	
	To simplify notations, we ignore the optimal strategy of $v_{*}^{i}(s,\pi_{*}^{1},\pi_{*}^{2})$ and instead represent the optimal accumulated expected reward  with the discounted criterion by $v_{*}^{i}(s)$ without ambiguity.
	It is not hard to find that
	\begin{align}
	v_{*}^{i}(s)=\max_{a_{1}}\min_{a_{2}}Q^{i}_{*}(s,a_{1},a_{2}).
	\end{align}
	Our goal is to characterize the strictly increasing property of optimal strategy $(\pi^{1}_{*},\pi^{2}_{*})$.
	We first prove that $Q^{i}_{i}(s,a_{1},a_{2})$ is strictly supermodular.
	Formally, we give the definition of the partially-ordered set as follows:
	\begin{definition}
		The set $X=\{y=(y_{1},y_{2},..., y_{n}): y_{i}\in R^{1}$ for $i=1,2,...,n\}$ with the ordering relation $\succ$ where $y^{'}\succ y^{''}$ in $X$ if $y_{i}^{'}>y_{i}^{''}$ in $R^{1}$ for $i=1,2,...,n$, is a partially order set.
		A partially-ordered set $(X, \succ)$ is referred as  lattice iff for all $a, b \in X$,
		\begin{eqnarray*}
			\begin{aligned}
				a \vee b & \triangleq \inf \{c \in X,| c \succ a, c \succ b \}\in X,\\
				a \wedge b & \triangleq \sup \{c \in X,| a \succ c, b \succ c\}\in X .
			\end{aligned}
		\end{eqnarray*}
	\end{definition}
	Here, operators $\vee$ and $\wedge$ are referred to $join$ and $meet$, respectively. Note that, $a \vee b$ stands for the smallest upper bound for $\{a, b\}$. Similarly, $a \wedge b$ represents the greatest lower bound for $\{a, b\}$ in the sense that $a \succ a \wedge b, b \succ a \wedge b$, and if $a \succ c$ and $b \succ c$, it is easily to acquire that $a \wedge b \succ c$.
	\begin{definition}
		Given any lattice $(X,\succ)$, a function $h: X \to \mathbb{R}$ is referred to be strictly supermodular if for all $a, b\in X$,
		\begin{equation}
		h(a \wedge b) +h(a \vee b) > h(a)+h(b).
		\end{equation}
		The function $h$ is referred to be strictly submodular if $-h$ is strictly supermodular.
	\end{definition}
	
	Note that when $X=X_{1}\times X_{2}$ is ordered coordinate-wise, supermodularity catches the idea of complementarity between $X_{1}$ and $X_{2}$ accurately. Actually, if we let $a=(x_{1},x_{2})$ and $b=(y_{1}, y_{2})$ with $x_{1} \succ y_{1}$  and  $y_{2} \succ x_{2}$, we have $a\vee b=(x_{1},y_{2})$ and $a \wedge b=(y_{1},x_{2})$. Then, we can write the inequality in the definition of strictly supermodularity as
	\begin{equation}
	h(x_{1},y_{2})-h(x_{1},x_{2}) >h(y_{1},y_{2})-h(y_{1},x_{2}).
	\end{equation}
	We present the main result for the individual decision problems, establishing the lattice structure of the optimal strategies and deriving monotonicity  of the solution to the complementary payoff parameters. Before giving the monotone structures of optimal strategies, we provide the following lemma which is a key step in deriving monotone structure of optimal strategies.
	\begin{lemma}
		For any lattices $(\mathcal{S}, \succ)$ and $(\mathcal{A}, \succ)$, let $Q_{*}: \mathcal{S} \times \mathcal{A} \to \mathbb{R}$ be a strictly supermodular function (with coordinate-wise order) and define
		\begin{equation}
		a^{*}(s)=\arg\max_{x\in \mathcal{A}_{s}} Q(s,a).
		\end{equation}
		If $s \succ s'$ and $a(s)\succ a(s')$, then $a^{*}(s)\succ a^{*}(s')$.
	\end{lemma}
	\begin{IEEEproof}
		See appendix B.
	\end{IEEEproof}
	Denote by $\overrightarrow{a_{1}}=(a_{1}^{+},a_{2}^{+})$, $\overrightarrow{a_{2}}=(a_{1}^{-},a_{2}^{-})$, $s'=(m+1, g^{'}_{s},g^{'}_{a})$, and $s=(m, g_{s},g_{a})$ for any $a_{1}^{+} > a_{1}^{-}$, $a_{2}^{+} > a_{2}^{-}$, where $a_{1}^{+},  a_{1}^{-}\in\mathcal{P}_{\mathcal{A}}
	$, $a_{2}^{+},a_{2}^{-}\in\mathcal{P}_{\mathcal{S}}$, $m\in \mathbb{Z}$, $g^{'}_{s}> g_{s}$, and $g^{'}_{a}> g_{a}$. It follows  that $s' \succ s$. We take $x=(s',\overrightarrow{a_{1}})$, $y=(s,\overrightarrow{a_{2}})$ and it follows  that  $x \succ y$.
	Note that in order to prove $Q_{*}(s,a_{1},a_{2})$ is strictly supermodular, it suffices to show that
	\begin{align*}
	Q_{*}^{i}&(x\vee y)+Q_{*}^{i}(x \wedge y) - Q_{*}^{i}(x)-Q_{*}^{i}(y) \\
	&=Q_{*}^{i}(s',\overrightarrow{a_{1}})+Q_{*}^{i}(s,\overrightarrow{a_{2}})- Q_{*}^{i}(s',\overrightarrow{a_{2}})-Q_{*}^{i}(s,\overrightarrow{a_{1}})
	> 0.
\end{align*}
	Define \begin{equation}
	\varepsilon\triangleq \frac{u(g_{a})u(g_{s})[q(a^{+}_{1},a^{+}_{2},g_{s},g_{a})-q(a^{-}_{1},a^{-}_{2},g_{s},g_{a})]}{u(g'_{a})u(g'_{s})[q(a^{+}_{1},a^{+}_{2},g'_{s},g'_{a})-q(a^{-}_{1},a^{-}_{2},g'_{s},g'_{a})]}
	\end{equation} in which $g_{s}$, $g_{a}$, $ g'_{s}$, and $g'_{a}$ belong to a finite set $\Xi$ and $a^{-}_{1}, a^{-}_{2}, a^{+}_{1}, a^{+}_{2}$ are the action power selected from the finite set $\mathcal{P}_{\mathcal{A}}$ and $\mathcal{P}_{\mathcal{S}}$ by the attacker and sensor, respectively.
	Due to the fact that $\varepsilon$ belongs to a finite set, there exists an upper bound which is denoted by $\varepsilon_{max}$. Therefore,  a sufficient condition for the strictly supermodularity
	of the $Q_{*}(s,a_{1},a_{2})$ function is acquired, that is, $\frac{v^{i}_{*}(0)-v^{i}_{*}(m+2)}{v^{i}_{*}(0)-v^{i}_{*}(m+1)}> \varepsilon_{max}$.
	\begin{definition}
		(strictly increasing stationary strategies) For a stationary strategy taken by the attacker $\{u(\cdot|s), s\in \mathcal{S}\}$, denote by $a_{1}(\pi^{1}=u|s)$ the action point of strategy $u$ in state $s$. This strategy is strictly increasing if $a_{1}(\pi^{1}=u|s=s_{2}) < a_{1}(\pi^{1}=u|s=s_{1})$ holds for any $s_{1}, s_{2}\in \mathcal{S}$ with $s_{1}\succ  s_{2}$.
	\end{definition}
	
	Notice that similar definition is applicable to the stationary strategy of the sensor.  The special structure of the optimal stationary strategy for the attacker-sensor game $\Gamma$ is established in the following.

		\begin{theorem}	
			If there exists a state $i$ such that $$\frac{v^{i}_{*}(0)-v^{i}_{*}(m+2)}{v^{i}_{*}(0)-v^{i}_{*}(m+1)} > \frac{r(0)+...+r(i+1)}{r(0)+...+r(i)} > \varepsilon_{max}$$ for the given $\varepsilon_{max}$ and $a_{2}^{+}a_{1}^{-}\geq a_{2}^{-}a_{1}^{+}$ holds for any $a_{1}^{+},a_{1}^{-}\in \mathcal{P}_{\mathcal{A}}$, $a_{2}^{+},a_{2}^{-}\in \mathcal{P}_{\mathcal{S}}$, where $a_{1}^{+} > a_{1}^{-}$ and $a_{2}^{+}>  a_{2}^{-}$.
			Then  any of the optimal stationary strategies $(\pi^{1}_{*},\pi^{2}_{*})$ of the attacker-sensor game is  strictly increasing for $s \succ i$.
		\end{theorem}
		\begin{IEEEproof}
			See appendix C.
	\end{IEEEproof}
	
	What we have assumed so far is that everything in the game is common knowledge for sensors and  attackers. But in practice there is some private information such as payoffs, type or preferences that is not known by the opposite player.
	So the next thing we are going to consider is the incomplete information scenarios.
	\section{Incomplete Information-Bayesian Game Framework}\label{sec:incomplete-information-game}
	\subsection{Game Formulation}
	
	The assumption that the sensor and attacker know the channel information of each other is hard to implement in a real scenario.
	In light of this, we extend the DoS attack to a Bayesian game model, where both the sensor and attacker have incomplete information\cite{Liang2020},  in other words, the sensor has its own channel gain $g_{k}^{s}$, but does not acquire the channel gains of the attacker $g_{k}^{a}$. Similar assumption applies to the attacker.
	
	Apart from the actual players such as  sensors and attackers in the game, we assume that  a special player called \textit{Nature} is existed.  The nature appoints a  random variable to the sensor and the attacker which represents the value of  type for the attacker (or the sensor) and provide the probability distribution for those types. In other words, Nature can randomly select a type for the attacker (or the sensor) according to the probability distribution of  types.
	
	The formal definitions of Bayesian game is provided as follows:
	\subsection{Bayesian Game  Framework}
	The Bayesian Game, denote by $\mathcal{G}$, is defined as  $<\mathcal{I}^{B},\mathcal{A}^{B},\Theta,\varDelta(\Theta),R^{B}>$ where
	
	\textbf{\textit{Players}}: $\mathcal{I}^{B}=\{I^{1}, I^{2}\}$ represents the set of players where $I^{1}$ stands for the attacker and $I^{2}$ denotes the sensor. Each player acts as a selfish and rational decision maker (this means a player always choose the action that brings the best response).
	
	\textbf{\textit{Actions}}: $\mathcal{A}^{B}=\{\mathcal{A}^{1},\mathcal{A}^{2}\}$ stands for  the action pair set. Denote  $\mathcal{A}^{i}$  as the action  set  chosen by  the  player $i$. The attacker decides the attacking energy $a_{k}\in \mathbb{P}_{\mathcal{A}}\triangleq [\mathcal{P}_{a_{1}},\cdots, \mathcal{P}_{a_{max}}]$, and the sensor adopts  the transmission energy $b_{k}\in\mathbb{P}_{\mathcal{S}}\triangleq [\mathcal{P}_{s_{1}},\cdots, \mathcal{P}_{s_{max}}].$  The action pair  is denoted by $(a_{k},b_{k})$.

	\textbf{\textit{Types}}: $\Theta=\{\Theta_{1},\Theta_{2}\}$ is the set of types for the sensor and attacker. The \textit{type} of a player is a kind of  private information which is not known by opponents but plays a role in the player's decision making. In this game, the channel gain is private, and hence it is known as the type of the sensor and the attacker, respectively, i.e., $\Theta_{i}=\Xi$ for $i=1,2$.
	
	\textbf{\textit{Belief}}: $\varDelta(\Theta)=(\varDelta(\Theta_{1}),\varDelta(\Theta_{2}))$ is joint probability distribution over the type of sensors and attackers. This belief is regarded as  the common knowledge shared by all the players. And the players will obtain a belief from the opponent's type according to channel state transition probability matrix $\mathcal{T}$.
	
	\textbf{\textit{Rewards:}} Denote $r^{i}$ as the immediate payoff function of player $i$ with $r^{i}:\mathcal{S}\times \mathcal{A}^{1} \times \mathcal{A}^{2}\rightarrow R$.
	The  immediate payoff of the attacker is provided  as
	\begin{align}
	r^{1}(m,a,b)=Tr[h^{m}(\bar{P})]+\alpha_{s}b-\alpha_{a}a.
	\end{align}
	
	In this Bayesian games, a pure strategy is a function $\pi^{1}(g^{s}_{k})$ for the sensor and $\pi^{2}(g^{a}_{k})$ for the attacker which is given in \eqref{strategypi} and it assigns an action that sensors or attackers will select when a particular type is acquired. 

	Denote  $g_{-i}$ as  the type of its opponent and $\mathcal{P}_{i}$ as the set of functions $p_{i}:\mathcal{G}\longrightarrow \mathcal{P}_{i}$ for player $i$, where $i\in\{I^{1}, I^{2}\}$.
	
	As we know,  the channel gains transition probability matrix  and stationary distribution are common knowledge to the sensor and attacker.

	\begin{definition}
		Given a strategy $a_{i}(g_{i})$ and $a'_{i}(g_{i})$, the strategy profile $\{a^{*}_{i},a^{*}_{-i}\}$ is a pure-strategy Bayesian Nash equilibrium if for each participant $i\in \{I^{1}, I^{2}\}$ and every $g_{i}\in \Xi$,
		\begin{align*}
		a_{i}^{*}(g_{i})=\arg\max_{a_{i}^{'}}\sum_{g_{-i}}r_{i}(a^{'}_{i},a^{*}_{-i}(g_{-i})|g_{i},g_{-i}) Pr(g_{-i}|g_{i}).
		\end{align*}
	\end{definition}
	
	That is, no matter what the type implementation is, changing the strategy $p^{*}_{i}(g_{i})$ will not benefit the player. The extension to a best mixed strategy pair $\{\pi^{*}_{1},\pi^{*}_{2}\}$ is given as follows
	\begin{align*}
	\pi^{*}_{i}(g_{i})=\mathop{\arg\max}_{\pi^{'}_{i}}\sum_{g_{-i}}r_{i}(\pi^{'}_{i},\pi^{*}_{-i}(g_{-i})|g_{i},g_{-i}) Pr(g_{-i}|g_{i}).
	\end{align*}
	Firstly, the matrix form of zero-sum Bayesian games is derived.  For a participant with type size $|\Theta_{1}|$ and action set size $|\mathbb{P}_{\mathcal{A}}|$, strategy set has $|\Theta_{1}|^{|\mathbb{P}_{\mathcal{A}}|}$ entries. The joint probability and two players' strategies are used to calculate the sensor and attacker's payment. This led to a size $|\Theta_{1}|^{|\mathbb{P}_{\mathcal{A}}|}\times |\Theta_{2}|^{|\mathbb{P}_{\mathcal{S}}|}$ matrix game, the game can be calculated by linear programming. 

	\section{Numerical Examples}\label{sec:numerical-examples}
	The numerical examples are given in this section to demonstrate that the results in our work are correct.
	First, the system parameters are provided as
	$A=1.2$, $C=0.7$, $Q=R=0.8$, $\sigma^{2}=0.5$.
	In our simulations, the action sets of the attacker and the sensor are $\{1,6\}$ and  $\{2,5\}$, respectively.  And the channel gain for attackers and sensors  are selected from the set $\Xi=\{0.8,0.6\}$. Besides, the channel state transition probability $\mathcal{T}$ is given as
	$$
	\mathcal{T}=
	\begin{bmatrix}
	\Pi(0.8|0.8)& \Pi(0.6|0.8)  \\
	\Pi(0.8|0.6)  &\Pi(0.6|0.6) \end{bmatrix}=
	\begin{bmatrix}
	1/2 & 1/2  \\
	1/2 & 1/2 \end{bmatrix}.
	$$
	Note that $\mathcal{T}$ is irreducible as the graph formed by this matrix is strongly connected. Also, $\mathcal{T}$ is aperiodicity due to the fact that
		the  diagonal elements of $\mathcal{T}$ are all positive.
	Define the learning rate as
	$
	\alpha_{k}=\frac{10}{15+count(s,a_{1},b_{1})},
	$
	where $count(s,a_{1},b_{1})$ represents the number of  occurrence of the combination $(s,a_{1},b_{1})$.
	Note that the decay learning rate designed above guarantees that the conditions in \textit{Assumption 2} is satisfied.
	Therefore, the state which is rarely accessed and action combinations will put more weight  on the next learning process.
	
	\begin{figure}
		\centering
		\includegraphics[width=0.48\textwidth]{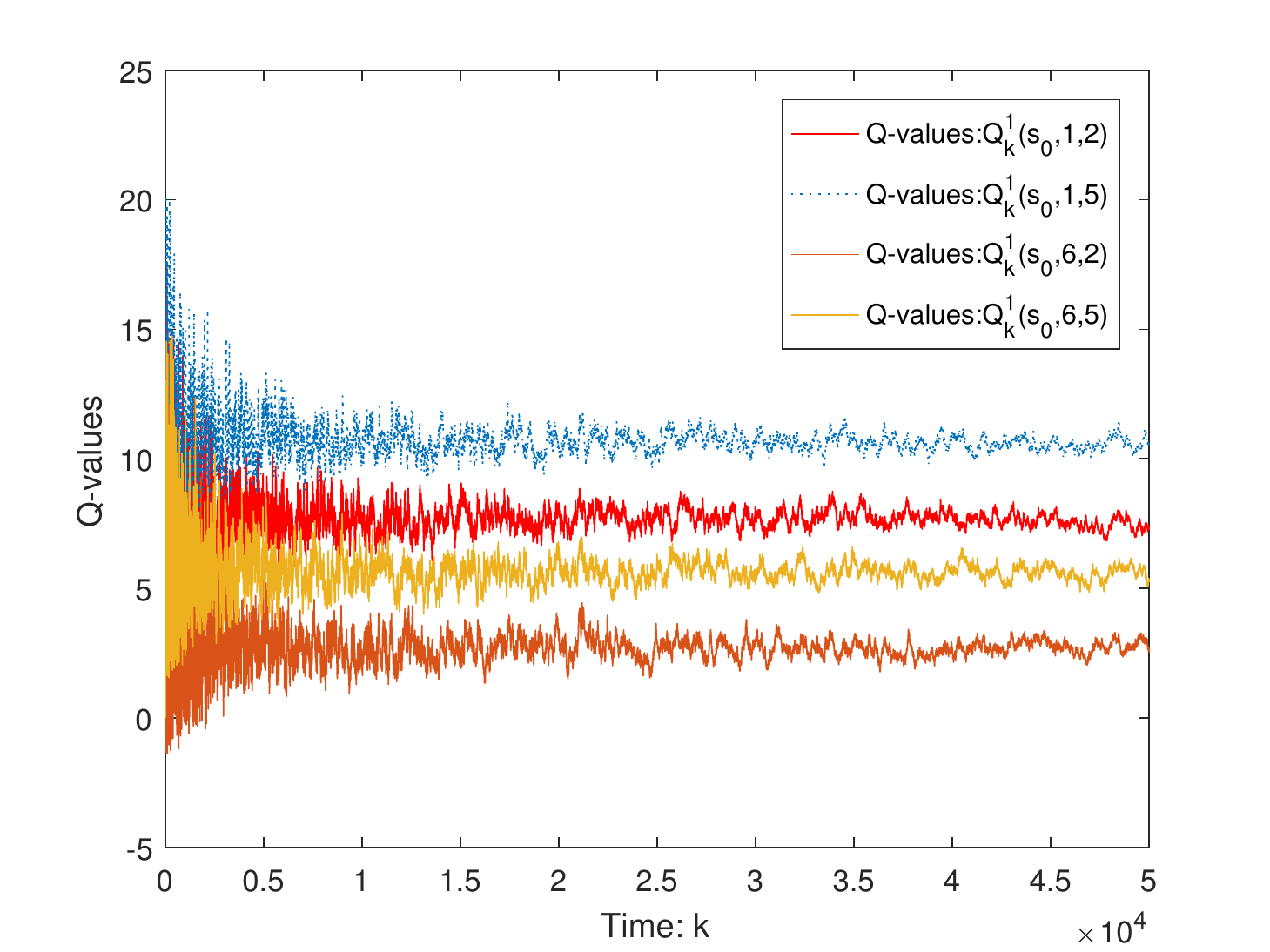}\\
		\caption{The Qvalues in the state of $s_{0}$ }
		\label{qvalue}
	\end{figure}

First, we present the calculation of \textit{Nash}$Q^{i}_{k}(s')$ which is the player $i$'s reward in state $s'$ for the chosen equilibria. It is easy to see that $\pi^{1}(s')\pi^{2}(s')Q^{i}_{k}(s')$ is a scalar. Obviously, we need to obtain a mixed-strategy Nash equilibrium denoted by $(\pi^{1}(s'),\pi^{2}(s'))$ for stage game $(Q^{1}_{k}(s'),Q^{2}_{k}(s'))$ with
the maximum value method or equivalent method.
We assume that the next state $s'=(0, 0.8,0.6)$ and the stage game is described in the Table I.
\begin{table}[!htbp]\scriptsize
	\centering
	\caption{Stage Game for state $s'$}\label{stagegame}
	\resizebox{0.49\textwidth}{8mm}{
		\begin{tabular}{|c|c|c|c|c|}
			\hline
			\multicolumn{2}{|c|}{ \multirow{2}*{Stage Game $(Q^{1}_{k}(s'),Q^{2}_{k}(s'))$}}& \multicolumn{2}{c|}{Attacker's action  $a_{1}$} \\
			\cline{3-4}
			\multicolumn{2}{|c|}{}&1&6\\
			\hline
			\multirow{2}*{Sensor's action $a_{2}$}&2&(-1.9906,1.9906)&(-4.9245,4.9245)\\
			\cline{2-4}
			&5&(3.0094,-3.0094)&(0.0755,-0.0755)\\
			\hline
		\end{tabular}
	}
\end{table}
Let $(p,1-p)$ be attacker's probability of taking action 1 and 6, and $(q,1-q)$ be sensor's probability of choosing energy level  2 and 5. Now we apply the Lemke-Howson method to calculate the mixed-strategy Nash equilibrium as
$\pi^{1}(s')=(0.4718, 0.5282)$ and
$\pi^{2}(s')=(0.2297, 0.7703)$, that is, in state $s'=(0, 0.8,0.6)$, $\mathbb{P}(a_{1}=1)=0.4718,\mathbb{P}(a_{1}=6)=0.5282$ and $\mathbb{P}(a_{2}=2)=0.2287,\mathbb{P}(a_{2}=5)=0.7703$.

	A learning player, such as the attacker, initializes $Q^{1}_{0}(s,a_{1},b_{1})=0$ and  $Q^{2}_{0}(s,a_{1},b_{1})=0$  for all $a_{1}\in \mathbb{P}_{\mathcal{A}}$,  $b_{1}\in \mathbb{P}_{\mathcal{S}},$ and $s\in \mathcal{S}$. A game begins with the initial state $s'=(0, 0.8,0.8)$. Players take their actions at the same time when they obtain their current state. Then, the next state, rewards of both the attacker and the sensor, and the energy level chosen by the opponent are obtained. Based on the rule  in \eqref{updaterule}, the Q-functions are updated by learning player. In the next state, the player repeats the procedure above. While at least one player gets its optimal $Q$-value, the procedure restarts. In the new stage, each player is randomly assigned a new state. The training stops after 50000 episodes.
	
	We can see that the result in Fig.~\ref{qvalue} is  coincident with the theoretical derivation. Our example proves that the learning Q-functions have the equilibrium strategies as $Q_{*}$. 	And the optimal Q-value of other state for every action pair is represented in Table II.
		\begin{table}[!htp]\scriptsize
		\centering
		\caption{Optimal Q-value of all state for every action pair}\label{stagegame}
		\resizebox{0.49\textwidth}{40mm}{
			\begin{tabular}{|c|c|c|c|c|}
				\hline
				\multicolumn{1}{|c|}{ \multirow{2}*{ $Q^{1}_{*}(s,a_{1},a_{2})$}}& \multicolumn{4}{c|}{Action pair $(a_{1},a_{2})$} \\
				\cline{2-5}
				&(1,2)&(1,5)&(6,2)&(6,5)\\
				\hline
				$s_{0}$&7.73&10.33&3.12&5.51\\
				\hline
				$s_{1}$&7.57&11.04&2.97&5.72\\
				\hline
				$s_{2}$&7.79&10.89&2.79&6.03\\
				\hline
				$s_{3}$&7.87&10.95&3.18&5.79\\
				\hline
				
				$s_{4}$&8.08&10.96&3.79&6.91\\
				\hline
				
				$s_{5}$&8.38&10.87&3.76&6.71\\
				\hline
				
				$s_{6}$&9.03&10.83&4.33&6.82\\
				\hline
				
				$s_{7}$&8.73&10.98&4.24&6.80\\
				\hline
				
				$s_{8}$&10.50&11.54&6.63&9.24\\
				\hline
				
				$s_{9}$&10.28&11.73&6.75&9.29\\
				\hline
				
				$s_{10}$&10.57&11.85&6.61&9.22\\
				\hline
				
				$s_{11}$&10.33&12.00&6.78&8.76\\
				\hline
				
				$s_{12}$&14.19&13.69&11.52&13.64\\
				\hline
				
				$s_{13}$&14.40&13.61&11.86&13.78\\
				\hline
				
				$s_{14}$&15.53&15.54&11.82&14.28\\
				\hline
				
				$s_{15}$&15.25&15.08&12.23&14.40\\
				\hline
				
				$s_{16}$&23.91&19.70&23.16&24.00\\
				\hline
				
				$s_{17}$&23.11&19.71&22.77&24.14\\
				\hline
				
				$s_{18}$&25.08&22.09&23.14&25.19\\
				\hline
				
				$s_{19}$&24.60&22.68&23.57&25.46\\
				\hline
			\end{tabular}
		}
	\end{table}\vspace{-3mm}

\begin{figure}
		\centering
		\includegraphics[width=0.48\textwidth]{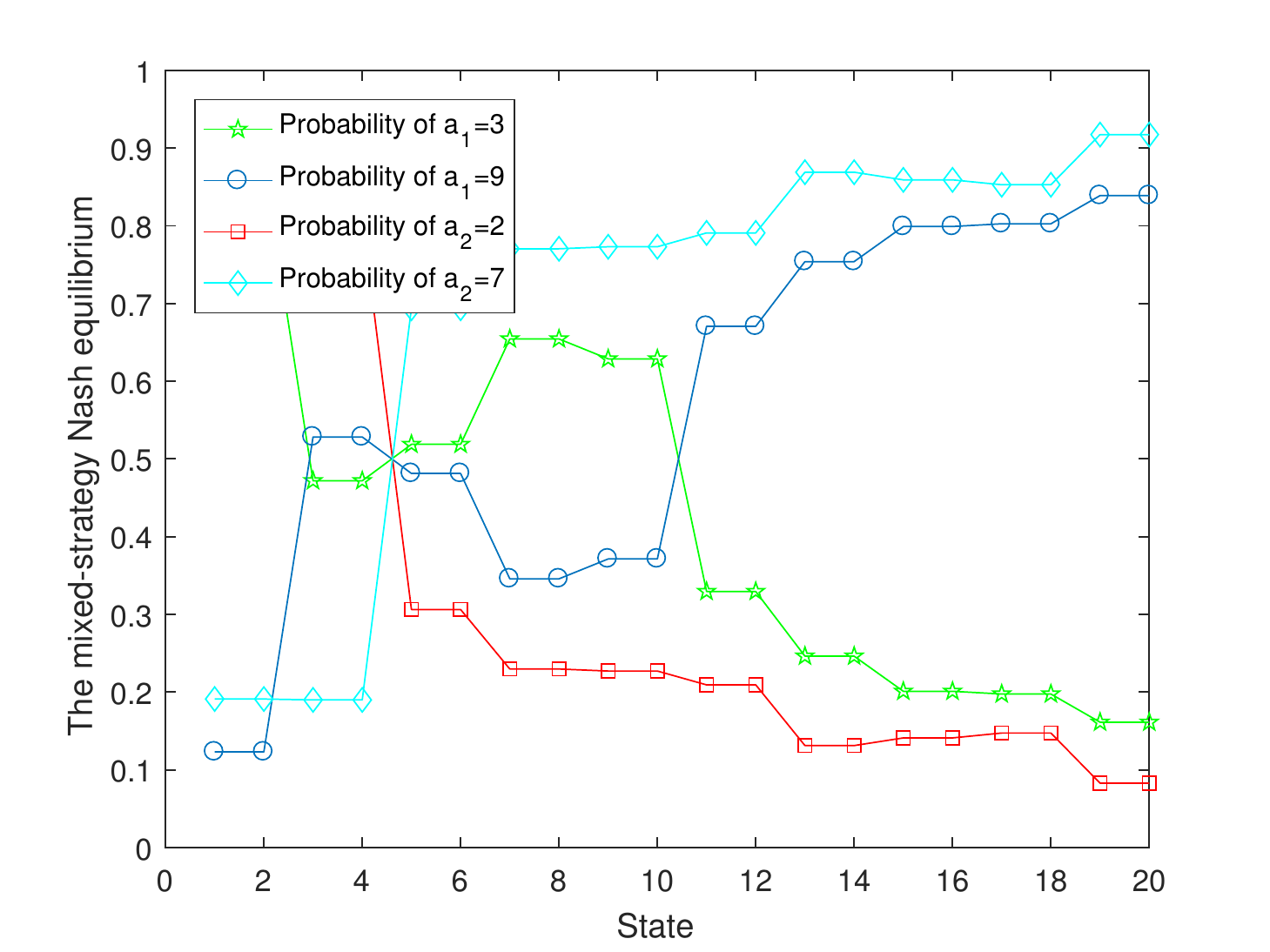}\\
		\caption{The mixed-strategy stationary Nash equilibrium for the attacker/sensor with different states.}
		\label{mon}
	\end{figure}

	Next,  we present that the optimal Nash stationary policies are monotone functions of state when the conditions are satisfied.  We assume that the system parameters are same as the previous scenario other than adjusting the energy level sets of the attacker and the sensor to $\{9, 3\}$ and $\{7, 2\}$,  respectively.
		It is easy to obtain  that
		$\frac{g_{s}a^{+}_{2}}{g_{a}a^{+}_{1}+\sigma^{2}}>\frac{g_{s}a^{-}_{2}}{g_{a}a^{-}_{1}+\sigma^{2}}, \forall{g_{s}, g_{a}\in\Xi},$
		where $\sigma^{2}=0.5$, $a^{+}_{1}=9$, $a^{-}_{1}=3$, $a^{+}_{2}=7$, $a^{-}_{2}=2$.
		Then, we have  $q(a^{+}_{1},a^{+}_{2},g_{s},g_{a})-q(a^{-}_{1},a^{-}_{2},g_{s},g_{a}) > 0$.  As depicted in Fig.~\ref{mon}, the optimal transmission schedule  for the sensor is transmitting the data packet in a  minimum energy level $2$ in a high probability when states $s\prec s_{5}=(1,0.8,0.8)$ and use power $7$ with the probability of $0.8323$  for state $s\succ s_{12}=(3,0.8,0.8)$; while the attacker adopts jamming power $3$ with a high probability when $s \prec s_{12}=(3,0.8,0.8)$ and $9$ for states $s \succ s_{12}=(3,0.8,0.8)$ with the probability of $0.9108$. This demonstrates that the optimal stationary strategy is monotone functions of state, which represents  that the higher the current states, the larger  the power actions of the sensor and the attacker.

	Finally, we present Bayesian Nash equilibria in incomplete information games in which  players only know their channel types.
		Based on the channel gains transition probability matrix $\mathcal{T}$, the stationary distribution of gains is obtained as $\mu=[0.5,0.5]$. Applying the result from Section IV, the mixed strategies $\pi^{*}_{1}(g_{a})$ and $\pi^{*}_{2}(g_{s})$ at the Bayesian Nash equilibrium is presented in the following.
		\begin{table}[!htp]
			
			
			\caption{The attacker's strategy $\pi^{*}_{1}(g_{a})$}
			\centering
			\label{tab:1}       
			
			
			\begin{tabular}{lll}
				
				\hline\noalign{\smallskip}
				
				& $g_{a}=0.6$ &$g_{a}=0.8$  \\
				
				\noalign{\smallskip}\hline\noalign{\smallskip}
				
				$Pr(a_{1}=1|g_{a})$ & 0.4058 & 0.0350 \\
				
				$Pr(a_{1}=6|g_{a}) $& 0.5942 & 0.9650 \\
				
				\noalign{\smallskip}\hline
				
			\end{tabular}
			
		\end{table}
		\begin{table}[!htp]
			\centering
			
			\caption{The sensor's strategy $\pi^{*}_{2}(g_{s})$}
			
			\label{tab:2}       
			
			
			\begin{tabular}{lll}
				
				\hline\noalign{\smallskip}
				
				& $g_{s}=0.6$ &$g_{s}=0.8$  \\
				
				\noalign{\smallskip}\hline\noalign{\smallskip}
				
				$Pr(a_{2}=2|g_{s})$ &0.4445 & 0.1723 \\
				
				$Pr(a_{2}=5|g_{s}) $ & 0.5555 & 0.8277 \\
				
				\noalign{\smallskip}\hline
				
			\end{tabular}
		\end{table}
		Notice that, from Table \ref{tab:1}, \ref{tab:2},
		regardless of the state of the attacker's channel, when $g_{s}=0.8$ is observed, the probability that the sensor chooses to transmit high power is 0.8277. If the channel of the sensor is changed to $g_{s}=0.6$, it would use a low power level with a probability of $0.44445$. Similarly interpretation can be acquired from the attacker's energy selection strategy $\pi^{*}_{1}(g_{a})$ in Table \ref{tab:1}.
	\section{Conclusions}\label{sec:conclusion}
	In this paper, we have discussed a cyber-physical system security issue, where a smart attacker deploys DoS attacks to destroy the transmission channel through which a sensor transmits data packets to the remote center. The sensor using a higher power level can guarantee a lower packet dropout rates and then improves the system performance.  In contrast, the attacker sends a jamming data packet to destroy the signal transmitted by the sensor and then causes the performance degradation.  With the goals of  two players in infinite-time horizon, we describe the conflicting characteristic between the attacker and the sensor with a general-sum stochastic game. And the Nash Q-learning techniques are applied to find an optimal solution at a Nash equilibrium. Also the convergence analysis of the proposed algorithm in our game is provided. Besides, the monotone structure of the optimal stationary strategy is constructed under  a sufficient condition.  For the incomplete information scenario, the original stochastic game is extended to the Bayesian game.
	\bibliographystyle{IEEE}
	
	\begin{appendices}
		\section{}
		\textbf{Proof of \textit{Theorem 2:}}
		Before proceeding, we provide an important lemma which presents a pseudo-contraction mapping for the convergence analysis of Q-learning algorithm.
		Denote  $\mathcal{Q}$  as the set of all $Q$ functions.
		
		\begin{lemma}[\textit{cf. [19]}]\label{map}
			Assume that  the mapping $f_{k}: \mathcal{Q}\rightarrow \mathcal{Q}$ satisfies the following condition if there exists a number $0<\varrho<1$ and a sequence $\kappa_{k}$ converging to zero with probability 1 such that $||f_{k}Q^{i}_{k}-f_{k}Q^{i}_{*}||\leq \varrho||Q^{i}_{k}-Q^{i}_{*}||+\kappa_{k}$ for all $Q^{i}_{k}\in \mathcal{Q}$ and $Q^{i}_{*}=\mathbb{E}[f_{k}Q^{i}_{*}]$, then the update defined by
			$
			Q^{i}_{k+1}=(1-\alpha_{k})Q^{i}_{k}+\alpha_{k}[f_{k}Q^{i}_{k}]
			$
			converges to $Q^{i}_{*}$ with probability $1$ provided that $\alpha_{k}$ satisfies \textit{Assumption} 2.
		\end{lemma}
		
		For this attacker-sensor  zero-sum  game, the mapping  $f_{k}$ is defined as follows.  Let $Q=(Q^{1}_{k},Q^{2}_{k})$, where $Q^{i}_{k}\in\mathbb{Q}^{i}$ for $i=1,2$, and $\mathcal{Q}=\mathcal{Q}^{1}\times\mathcal{Q}^{2}$. $f_{k}:\mathcal{Q}\rightarrow\mathcal{Q}$ is a function from the  $\mathcal{Q}$ to $\mathcal{Q}$, $f_{k}\mathcal{Q}=(f_{k}\mathcal{Q}^{1},f_{k}\mathcal{Q}^{2})$, where
		\begin{align}\label{defmap}
			f_{k}Q^{i}_{k}(s,a_{1},a_{2})=r^{i}(s,a_{1},a_{2})+\beta \psi^{1}(s')\psi^{2}(s')Q^{i}_{k}(s')
		\end{align}
		for $i=1,2$, where $s'$ is the next state.
		
		Based on the \textit{lemma \ref{map}} and the equation \eqref{defmap}, we can obtain that
		$\mathbb{E}[f_{k}Q_{*}]=Q_{*}$ 	for the proposed two-player stochastic game, where $Q_{*}=(Q^{1}_{*},Q^{2}_{*})$. Since $v^{i}(s',\psi^{1}_{*},\psi^{2}_{*})$ is the player $i$'s Nash equilibrium reward for the state game $(Q^{1}_{*}(s'),Q^{2}_{*}(s'))$, and $(\psi^{1}_{*}(s),\psi^{2}_{*}(s))$ is its Nash equilibrium strategy, it follows $v^{i}(s',\psi^{1}_{*},\psi^{2}_{*})=\psi^{1}_{*}(s')\psi^{2}_{*}(s')Q^{i}_{*}(s')$. Thus, we obtain that
		\begin{equation*}
			\begin{aligned}
				&Q^{i}_{*}(s,a_{1},a_{2})\\
				&=r^{i}(s,a_{1},a_{2})
				+\beta\sum_{s^{'}\in \mathcal{S}}Pr(s^{'}|s,a_{1},a_{2})v^{i}(s^{'},\psi_{*}^{1},\psi_{*}^{2})
				\\
				&=\sum_{s^{'}\in \mathcal{S}}Pr(s^{'}|s,a_{1},a_{2})\Big( r^{i}(s,a_{1},a_{2})+\beta\psi_{*}^{1},\psi_{*}^{2}Q^{i}_{*}(s')\Big)\\
				&=\mathbb{E}[f_{k}Q^{i}_{*}(s,a_{1},a_{2})].
			\end{aligned}
		\end{equation*}  However, our convergence analysis requires that the stage games possess global optima, or alternatively, that they possess saddle points at every state $s$. Besides, they should  order the attacker or the sensor to select either global optima or saddle points to carry out the update of Q-values.
		\begin{figure*}
			\normalsize
			\setcounter{mytempeqncnt}{\value{equation}}
			\renewcommand{\theequation}{A\arabic{equation}}
			\setcounter{equation}{4}
			\begin{align*}
				\Delta_{2}&\triangleq q(a_{1}^{+},a_{2}^{+}, g'_{s},g'_{a})u(g'_{s})u(g'_{a})v_{*}^{i}(0)
				+(1-q(a_{1}^{+},a_{2}^{+}, g'_{s},g'_{a}))u(g'_{s})u(g'_{a})v_{*}^{i}(m+2)+q(a_{1}^{-},a_{2}^{-}, g_{s}, g_{a})u(g_{s})u(g_{a})v_{*}^{i}(0)+\nonumber\\
				&\,\,\,\,\,\,\,\,\,(1-q(a_{1}^{-},a_{2}^{-}, g_{s}, g_{a}))u(g_{s})u(g_{a})v_{*}^{i}(m+1)
				- q(a_{1}^{-},a_{2}^{-}, g_{s}, g_{a})u(g_{s})u(g_{a})v_{*}^{i}(0)
				-(1-q(a_{1}^{-},a_{2}^{-}, g_{s}, g_{a}))u(g_{s})u(g_{a}) \nonumber\\
				&\,\,\,\,\,\,\,\,\,v_{*}^{i}(m+1)-q(a_{1}^{+},a_{2}^{+}, g_{s}, g_{a})u(g_{s})u(g_{a})v_{*}^{i}(0)
				-(1-q(a_{1}^{+},a_{2}^{+}, g_{s}, g_{a}))u(g_{s})u(g_{a})v_{*}^{i}(m+1)
				\triangleq u(g'_{s})u(g'_{a})[q(a_{1}^{+},a_{2}^{+},\nonumber\\& \,\,\,\,\,\,\,\,\,g'_{s},g'_{a})-q(a_{1}^{-},a_{2}^{-}, g'_{s},g'_{a})][v_{*}^{i}(0)-v_{*}^{i}(m+2)]
				-u(g_{s})u(g_{a})[q(a_{1}^{+},a_{2}^{+}, g_{s}, g_{a})-q(a_{1}^{-},a_{2}^{-}, g_{s}, g_{a})][v_{*}^{i}(0)-v_{*}^{i}(m+1)].
			\end{align*}\vspace{-7mm}
			\begin{flalign*}
				&\Delta_{1}\triangleq r^{i}(m+1,a_{1}^{+},a_{2}^{+})+r^{i}(m,a_{1}^{-},a_{2}^{-})-r^{i}(m+1,a_{1}^{-},a_{2}^{-})-r^{i}(m,a_{1}^{+},a_{2}^{+})\triangleq Tr[h^{m+1}(\bar{P})]+\alpha_{s}a_{2}^{+}-\alpha_{a}a_{1}^{+}+Tr[h^{m}(\bar{P}\nonumber\\&\,\,\,\,\,\,\,\,\,\,\,\,\,\,\,\,)]+\alpha_{s}a_{2}^{-}-\alpha_{a}a_{1}^{-}-(Tr[h^{m+1}(\bar{P})]+\alpha_{s}a_{2}^{-}-\alpha_{a}a_{1}^{-})-(Tr[h^{m}(\bar{P})]+\alpha_{s}a_{2}^{+}-\alpha_{a}a_{1}^{+}).
			\end{flalign*}
			\setcounter{equation}{\value{mytempeqncnt}}
			\hrulefill 
			\vspace*{2pt} 
		\end{figure*}
		What we need to do is proving  that the $f_{k}$ mapping  is a pseudo-contraction mapping. That is, our $f_{k}$ satisfies $||f_{k}Q-f_{k}\widetilde{Q}||\leq \varrho ||Q-\widetilde{Q}||$ for all $Q, \widetilde{Q}\in \mathcal{Q}$. Before that,
		we define
		\begin{eqnarray*}
			\begin{aligned}
				||Q-\widetilde{Q}||&\triangleq \max_{i}\max_{s}||Q^{i}(s)-\widetilde{Q}^{i}_{s}||_{(i,s)}\\ &\triangleq \max_{i}\max_{s}\max_{a_{1},a_{2}}||Q^{i}(s,a_{1},a_{2})-\widetilde{Q}^{i}_{s,a_{1},a_{2}}||.
			\end{aligned}
		\end{eqnarray*}
		Now, the statement that mapping $f_{k}$ is a contraction mapping operator is  demonstrated, i.e.,  $||f_{k}Q-f_{k}\widetilde{Q}|| \leq \beta||Q-\widetilde{Q}||$.
		Note that
		\begin{equation*}
			\begin{aligned}||f_{k}Q&-f_{k}\widetilde{Q}||
				\\ &=\max_{i}\max_{s}|\beta \psi^{1}(s)\psi^{2}(s)Q^{i}(s)-\beta \widetilde{\psi}^{1}(s)\widetilde{\psi}^{2}(s)\widetilde{Q}^{i}(s)|
				\\ &= \max_{i} \beta  |\psi^{1}(s)\psi^{2}(s)Q^{i}(s)-\widetilde{\psi}^{1}(s)\widetilde{\psi}^{2}(s)\widetilde{Q}^{i}(s)|.
			\end{aligned}
		\end{equation*}
		We proceed to prove that
		\begin{align*}
			|\psi^{1}(s)\psi^{2}(s)Q^{i}(s)-\widetilde{\psi}^{1}(s)\widetilde{\psi}^{2}(s)\widetilde{Q}^{i}(s)|\leq ||Q^{i}(s)-\widetilde{Q}^{i}(s)||.
		\end{align*}
		For the sake of simplicity,   $\psi^{i}(s)\pi^{-i}(s)$  is  represented as  $\psi^{1}(s)\psi^{2}(s)$, and $\widetilde{\psi}^{i}(s)\widetilde{\psi}^{-i}(s)$ is rewritten as $\widetilde{\psi}^{1}(s)\widetilde{\psi}^{2}(s)$. The proposition we want to prove is
		\begin{align*}
			|\psi^{i}(s)\psi^{-i}(s)Q^{i}(s)-\widetilde{\psi}^{i}(s)\widetilde{\psi}^{-i}(s)\widetilde{Q}^{i}(s)|\leq ||Q^{i}(s)-\widetilde{Q}^{i}(s)||.
		\end{align*}
		Suppose that $(\psi^{i}(s), \psi^{-i}(s))$ and $(\widetilde{\psi}^{i}(s)\widetilde{\psi}^{-i}(s))$ satisfy the Assumption 4, which means they are global optimal points or saddle points.  If $\psi^{i}(s)\psi^{-i}(s)Q^{i}(s) \geq \widetilde{\psi}^{i}(s)\widetilde{\psi}^{-i}(s)\widetilde{Q}^{i}(s)$, we have
		\begin{equation*}\begin{aligned}
				&|\psi^{i}(s)\psi^{-i}(s)Q^{i}(s)-\widetilde{\psi}^{i}(s)\widetilde{\psi}^{-i}(s)\widetilde{Q}^{i}(s)|
				\leq 	\psi^{i}(s)\psi^{-i}(s)Q^{i}(s)\\&-\psi^{i}(s)\psi^{-i}(s)\widetilde{Q}^{i}(s) \leq || Q^{i}(s)-\widetilde{Q}^{i}(s) ||.
			\end{aligned}
		\end{equation*}
		If $\psi^{i}(s)\psi^{-i}(s)Q^{i}(s) \leq \widetilde{\psi}^{i}(s)\widetilde{\psi}^{-i}(s)\widetilde{Q}^{i}(s)$, then
		\begin{equation*}
			\begin{aligned}
				\widetilde{\psi}^{i}(s)\widetilde{\psi}^{-i}(s)\widetilde{Q}^{i}(&s)-\psi^{i}(s)\pi^{-i}(s)Q^{i}(s) \\ &\leq \widetilde{\psi}^{i}(s)\widetilde{\psi}^{-i}(s)Q^{i}(s)-\widetilde{\psi}^{i}(s)\widetilde{\psi}^{-i}(s)\widetilde{Q}^{i}(s),
			\end{aligned}
		\end{equation*} and the following  proof is  analogous to the above.
		Thus,
		\begin{equation*}
			\begin{aligned}
				||f_{k}Q&-f_{k}\widetilde{Q}||
				\\ &\leq \beta\max_{i}\max_{s}| \psi^{1}(s)\psi^{2}(s)Q^{i}(s)-\widetilde{\psi}^{1}(s)\widetilde{\psi}^{2}(s)\widetilde{Q}^{i}(s)|
				\\ & \leq \beta \max_{i}\max_{s}||Q^{i}(s)-\widetilde{Q}^{i}(s)||=\beta||Q-\hat{Q}||.
			\end{aligned}
		\end{equation*}
		The proof of \textit{Theorem 2} is mainly based on the \textit{Lemma A.1,} which presents  the  convergence analysis provided the two conditions are satisfied. Note that $f_{k}$ is a contraction operator, and we prove that $f_{k}$ is also a pseudo-contraction mapping. Furthermore,  the constant point condition, $\mathbb{E}(f_{k}Q_{*})=Q_{*}$, is obtained. Hence, the process
		$
		Q_{k+1}=(1-\alpha_{k})Q_{k}+\alpha_{k}[f_{k}Q_{k}]
		$
		converges to $Q_{*}$.\vspace{-2mm}
		\section{}
		\textbf{Proof of \textit{Lemma 3.1:}}
			%
			Assuming $s \succ s^{'}$ and $\mathcal{A}(s) \succ \mathcal{A}(s')$, take any $p\in a ^{*}(s)$ and $p' \in a^{*}(s')$. In order to show that $a^{*}(s) \succ a^{*}(s')$, we need to show that
			$p\vee p' \in a^{*}(s)$ and $p\wedge p' \in a^{*}(s')$. For this, it suffices to show that $p\vee p'\in \mathcal{A}(s), p\wedge p' \in \mathcal{A}(s')$, $Q_{*}(s, p\vee p')=Q_{*}(s,p)$ and $Q_{*}(s, p\wedge p')=Q_{*}(s',p')$. First, since $p\in a^{*}(s)\subset \mathcal{A}(s), p\in \mathcal{A}(s)$. Similarly,  $ p' \in \mathcal{A}(s')$.
			Since $\mathcal{A}(s) \succ \mathcal{A}(s')$, we obtain $p\vee p' \in \mathcal{A}(s)$ and $p\wedge p' \in \mathcal{A}(s')$. To show $Q_{*}(s, p\vee p')=Q_{*}(s, p)$ and $Q_{*}(s, p\wedge p')=Q_{*}(s', p')$, note that since $p \in a^{*}(s)$ and $p \vee p' \in \mathcal{A}(s)$, $Q_{*}(s, p \vee p')\leq Q_{*}(s, p)$. Similarly, $Q_{*}(s', p \wedge p')\leq Q_{*}(s', p')$. If either of these inequalities were strict, we would have
			\begin{equation*}
				Q_{*}(s, p\vee p')+Q_{*}(s', p \wedge p')< Q_{*}(s, p)+Q_{*}(s', p'),
			\end{equation*}
			contradicting the strictly supermodularity condition of $Q_{*}(s, a)$ above. Therefore, $Q_{*}(s, p\vee p')=Q_{*}(s,p)$ and $Q_{*}(s', p \wedge p')=Q_{*}(s', p')$.
			
			Note that when the optimization function $Q_{*}(s,a)$ is a strictly supermodular over a lattice, the above result implies that the optimal solutions are strictly increasing.
		
		Next, we establish the monotonicity of $(\pi^{1}_{*},\pi^{2}_{*})$ via checking a sufficient condition that $Q_{*}(s,a)$ is a strictly supermodular function as mentioned previously. \vspace{-2mm}
		\section{}
			\textbf{Proof of \textit{Theorem 3:}}
			First, assume that $(\pi^{1}_{*},\pi^{2}_{*})$ is the optimal  stationary strategy for  the attacker-sensor game. In order to prove the strictly increasing of $\pi^{1}_{*}$ and $\pi^{2}_{*}$, it is sufficient to verify that $Q^{i}_{*}(s,a_{1},a_{2})$ is strictly supermodular.
			It suffices to prove that  $Q_{*}^{i}(s',\overrightarrow{a_{1}})+Q_{*}^{i}(s,\overrightarrow{a_{2}})- Q_{*}^{i}(s',\overrightarrow{a_{2}})-Q_{*}^{i}(s,\overrightarrow{a_{1}})
			> 0$ for $\overrightarrow{a_{1}} \succ \overrightarrow{a_{2}}$ and $s'=(m+1,g'_{a},g'_{s}) \succ s=(m,g_{a},g_{s})$, which  is equivalent to verify that $\Delta_{1}+\beta\Delta_{2}> 0 $, as shown in the top of this page, for any $\overrightarrow{a_{1}} \succ \overrightarrow{a_{2}}$ and $s, s'\in \mathcal{S}$.  One easily obtains that $\Delta_{1}=0$. Since $\beta>0$, it is sufficient to verify that $\Delta_{2} >0$.
			First, we have that $$\frac{v^{i}_{*}(0)-v^{i}_{*}(m+2)}{v^{i}_{*}(0)-v^{i}_{*}(m+1)}  > \varepsilon_{max}$$ and $a_{2}^{+}a_{1}^{-} > a_{2}^{-}a_{1}^{+}$ holds for any $a_{1}^{+},a_{1}^{-}\in \mathcal{P}_{\mathcal{A}}$, $a_{2}^{+},a_{2}^{-}\in \mathcal{P}_{\mathcal{S}}$ and  $a_{1}^{+}> a_{1}^{-}$, $a_{2}^{+}> a_{2}^{-}$. It follows that
			$\frac{g_{s}^{'}a_{2}^{+}}{g_{a}^{'}a_{1}^{+}+\sigma^{2}}> \frac{g_{s}^{'}a_{2}^{-}}{g_{a}^{'}a_{1}^{-}+\sigma^{2}}$. This is equivalent to that the SINR of the first transmission is larger than the second one.
			Notice that the symbol error rate (SER) is tightly connected with the SINR as $
			SER=2Q(\sqrt{\alpha SINR}),$
			where $\alpha>0$ is a parameter and $Q(x)\triangleq \frac{1}{\sqrt{2\pi}}\int_{x}^{\infty}exp(-\eta^{2}/2)d\eta$ is nonincreasing function. Thus, the symbol error rate (SER) of the first transmission is lower than the second one. We then obtain that the reception rate $q(a^{+}_{1},a^{+}_{2},g_{s},g_{a})-q(a^{-}_{1},a^{-}_{2},g_{s},g_{a})> 0$ for any channel gain  $g_{s}$ and $g_{a}$. Define
			\begin{equation*}
				\begin{aligned}
					\chi &=\frac{v^{i}_{*}(0)-v^{i}_{*}(m+2)}{v^{i}_{*}(0)-v^{i}_{*}(m+1)},\\
					\phi &= \frac{u(g_{a})u(g_{s})[q(a^{+}_{1},a^{+}_{2},g_{s},g_{a})-q(a^{-}_{1},a^{-}_{2},g_{s},g_{a})]}{u(g'_{a})u(g'_{s})[q(a^{+}_{1},a^{+}_{2},g'_{s},g'_{a})-q(a^{-}_{1},a^{-}_{2},g'_{s},g'_{a})]}.
				\end{aligned}
			\end{equation*}
			It follows that $\chi \geq \phi $. Hence, we  can obtain that
			\begin{align}
				\Delta_{2}&=u(g'_{s})u(g'_{a})[q(a_{1}^{+},a_{2}^{+}, g'_{s},g'_{a})-q(a_{1}^{-},a_{2}^{-}, g'_{s},g'_{a})][v_{*}^{i}(0) \nonumber\\
				&-v_{*}^{i}(m+2)]
				-u(g_{s})u(g_{a})[q(a_{1}^{+},a_{2}^{+}, g_{s}, g_{a})-q(a_{1}^{-},a_{2}^{-},\nonumber\\& g_{s}, g_{a})][v_{*}^{i}(0)-v_{*}^{i}(m+1)]>0.	\end{align}
			Now since $Q_{*}^{i}(s',\overrightarrow{a_{1}})+Q_{*}^{i}(s,\overrightarrow{a_{2}})-Q_{*}^{i}(s',\overrightarrow{a_{2}})-Q_{*}^{i}(s,\overrightarrow{a_{1}})=\Delta_{1}+\beta\Delta_{2}> 0$ for any  $\overrightarrow{a_{1}} \succ \overrightarrow{a_{2}}$ and $s' \succ s$, it follows that $Q^{i}_{*}(s,a_{1},a_{2})$ is strictly supermodular. According to the \textit{Lemma 3.1}, we can obtain that $\arg\max_{s\in \mathcal{S}}Q^{i}_{*}(s,a_{1},a_{2})$ is strictly increasing in $(a_{1},a_{2})$. Due to the fact that $v_{*}^{i}(s)=\max_{a_{1}}\min_{a_{2}}Q^{i}_{*}(s,a_{1},a_{2})$, we can obtain that the energy action of attacker $a_{1}(s)$ is strictly increasing with the state $s$ when fixing the energy action of sensor $a_{2}$ in the \textit{Nash} equilibrium points. The same things happen to the sensor. Thus, the sufficient condition for the strictly increasing structure of the optimal strategies $(\pi_{*}^{1},\pi_{*}^{2})$ of the attacker-sensor game is obtained.
	\end{appendices}
	
\end{document}